\documentclass[final]{agujournal2019}
\usepackage{url} %this package should fix any errors with URLs in refs.
\usepackage{lineno}
\usepackage{amsmath}
\usepackage{color}
\usepackage{multirow}

\draftfalse

\journalname{JGR: Space Physics}

\begin{document}

%% ------------------------------------------------------------------------ %%
%  Title
%
% (A title should be specific, informative, and brief. Use
% abbreviations only if they are defined in the abstract. Titles that
% start with general keywords then specific terms are optimized in
% searches)
%
%% ------------------------------------------------------------------------ %%

% Example: \title{This is a test title}

\title{A gray-box model for a probabilistic estimate of regional ground magnetic perturbations: Enhancing the NOAA operational Geospace model with machine learning}

%% ------------------------------------------------------------------------ %%
%
%  AUTHORS AND AFFILIATIONS
%
%% ------------------------------------------------------------------------ %%

% Authors are individuals who have significantly contributed to the
% research and preparation of the article. Group authors are allowed, if
% each author in the group is separately identified in an appendix.)

% List authors by first name or initial followed by last name and
% separated by commas. Use \affil{} to number affiliations, and
% \thanks{} for author notes.
% Additional author notes should be indicated with \thanks{} (for
% example, for current addresses).

% Example: \authors{A. B. Author\affil{1}\thanks{Current address, Antartica}, B. C. Author\affil{2,3}, and D. E.
% Author\affil{3,4}\thanks{Also funded by Monsanto.}}

\authors{E. Camporeale\affil{1,2}, M.~D. Cash\affil{3}, H.~J. Singer\affil{3}, C.~C. Balch\affil{3}, Z. Huang\affil{4},~ G. Toth\affil{4}}

 \affiliation{1}{CIRES, University of Colorado, Boulder, CO, USA}
 \affiliation{2}{Center for Mathematics and Computer Science (CWI), Amsterdam, Netherlands}
 \affiliation{3}{NOAA, Space Weather Prediction Center, Boulder, CO 80305}
 \affiliation{4}{Department of Climate and Space Sciences and Engineering, University of Michigan, Ann Arbor, MI, USA}

%\affiliation{=number=}{=Affiliation Address=}
%(repeat as many times as is necessary)

%% Corresponding Author:
% Corresponding author mailing address and e-mail address:

% (include name and email addresses of the corresponding author.  More
% than one corresponding author is allowed in this LaTeX file and for
% publication; but only one corresponding author is allowed in our
% editorial system.)

% Example: \correspondingauthor{First and Last Name}{email@address.edu}

\correspondingauthor{Enrico Camporeale}{enrico.camporeale@noaa.gov}

%% Keypoints, final entry on title page.

%  List up to three key points (at least one is required)
%  Key Points summarize the main points and conclusions of the article
%  Each must be 100 characters or less with no special characters or punctuation and must be complete sentences

% Example:
% \begin{keypoints}
% \item	List up to three key points (at least one is required)
% \item	Key Points summarize the main points and conclusions of the article
% \item	Each must be 100 characters or less with no special characters or punctuation and must be complete sentences
% \end{keypoints}

\begin{keypoints}
\item We present a new model to forecast the maximum value of $dB/dt$ over 20-minute intervals at specific locations
\item The model provides a probabilistic forecast of exceeding a pre-defined threshold at a given location
\item The ML-enhanced algorithm consistently improves the predictive metrics of the physics-based model

\end{keypoints}

%% ------------------------------------------------------------------------ %%
%
%  ABSTRACT and PLAIN LANGUAGE SUMMARY
%
% A good Abstract will begin with a short description of the problem
% being addressed, briefly describe the new data or analyses, then
% briefly states the main conclusion(s) and how they are supported and
% uncertainties.

% The Plain Language Summary should be written for a broad audience,
% including journalists and the science-interested public, that will not have 
% a background in your field.
%
% A Plain Language Summary is required in GRL, JGR: Planets, JGR: Biogeosciences,
% JGR: Oceans, G-Cubed, Reviews of Geophysics, and JAMES.
% see http://sharingscience.agu.org/creating-plain-language-summary/)
%
%% ------------------------------------------------------------------------ %%

%% \begin{abstract} starts the second page

\begin{abstract}
We present a novel algorithm that predicts the probability that the time derivative of the horizontal component of the ground magnetic field $dB/dt$ exceeds a specified threshold at a given location. This quantity provides important information that is physically relevant to Geomagnetically Induced Currents (GIC), which are electric currents { associated to} sudden changes in the Earth's magnetic field due to Space Weather events. The model follows a 'gray-box' approach by combining the output of a physics-based model with machine learning. Specifically, we combine the University of Michigan's Geospace model that is operational at the NOAA Space Weather Prediction Center, with a boosted ensemble of classification trees.  We discuss the problem of re-calibrating the output of the decision tree to obtain reliable probabilities. The performance of the model is assessed by typical metrics for probabilistic forecasts: Probability of Detection and False Detection, True Skill Statistic, Heidke Skill Score, and Receiver Operating Characteristic curve.
{We show that the ML enhanced algorithm consistently improves all the metrics considered.}
\end{abstract}

%\section*{Plain Language Summary}
%[ enter your Plain Language Summary here or delete this section]

%% ------------------------------------------------------------------------ %%
%
%  TEXT
%
%% ------------------------------------------------------------------------ %%

%%% Suggested section heads:
% \section{Introduction}
%
% The main text should start with an introduction. Except for short
% manuscripts (such as comments and replies), the text should be divided
% into sections, each with its own heading.

% Headings should be sentence fragments and do not begin with a
% lowercase letter or number. Examples of good headings are:

% \section{Materials and Methods}
% Here is text on Materials and Methods.
%
% \subsection{A descriptive heading about methods}
% More about Methods.
%
% \section{Data} (Or section title might be a descriptive heading about data)
%
% \section{Results} (Or section title might be a descriptive heading about the
% results)
%
% \section{Conclusions}

\section{Introduction}
Geomagnetically induced currents (GIC) represent one of the most severe risks posed by space weather events on our infrastructure on the ground, such as high-voltage power transmission systems.  GICs are caused by sudden variations of the Earth's magnetic field that, through Faraday's law, induce a variation of the electric field \cite{boteler1998, pirjola2000, lanzerotti2001, pulkkinen2005, pirjola2007, schrijver2013}.
The induced electric fields responsible for GICs can be estimated from the amplitude of the time derivative of magnetic fluctuations, often denoted as $dB/dt$, when combined with information of local earth conductivity characteristics \cite{boteler1998b, pirjola2002,viljanen2004,ngwira2008,horton2012}. Hence, much attention has been dedicated to understanding and forecasting $dB/dt$ \cite{viljanen1997,viljanen2001}.\\
Previous works on forecasting $dB/dt$ can generally be divided into empirical and physics-based models. Empirical models exploit the statistical relationships between input quantities, such as solar wind observations recorded by satellites orbiting around L1 (first Lagrangian point) and the observed 
$dB/dt$ at a specific station, with a typical time-lag ranging between 15 and 60 minutes. Those statistical relationship can then be encoded into a regression model, in the form of a neural network, or a linear filter. Empirical models include \citeA{gleisner2001, weigel2002,weigel2003,wintoft2005,wintoft2005b,weimer2013,wintoft2015, lotz2015}.\\
On the other hand, physics-based models follow the evolution in time and space of the plasma and the electromagnetic field surrounding Earth and derive the ground magnetic field perturbation from physical laws. Typically the spatial domain is divided in sub-regions, where the MHD approximation is used in the outer magnetosphere, while the inner magnetosphere and the transition to ionosphere are modeled by including kinetic processes. Examples of physics-based models that can, in principle, forecast $dB/dt$ given the conditions of the solar wind observed at L1 are OpenGGCM (Open General Geospace Circulation Model, \citeA{raeder1998}), GAMERA (Grid Agnostic MHD for Extended Research Applications, \citeA{zhang2019}), and SWMF (Space Weather Modeling Framework, \citeA{toth2005}). Several works have assessed the ability of physics-based models to forecast geomagnetic perturbations and more generally to recover plasma and field conditions as observed in the data (see, e.g. \citeA{yu2008,welling2010,pulkkinen2011,rastatter2011,rastatter2013,gordeev2015,jordanova2018,welling2019}).
The validation and comparisons of different models for predicting $dB/dt$ was specifically tackled in \citeA{pulkkinen2013} in order to support selecting a model to transition to operations at NOAA's Space Weather Prediction Center (SWPC). As a result of that comparison, the University of Michigan's SWMF model, henceforth referred to as the Geospace model, was selected for transition to real-time operations.\\

In this paper we present a new model for predicting whether $dB/dt$ will exceed given thresholds in a given time interval at specific locations. The model builds on the physics-based Geospace model. We show that the skill of the physics-based model can be considerably enhanced with a machine learning technique, improving all the performance metrics considered.

\subsection{The Geospace model at NOAA/SWPC}
The Geospace model that runs operationally at NOAA/SWPC is a version of the Space Weather Modeling Framework developed by the University of Michigan \cite{toth2005,toth2012}, that couples the following three physics domains. The outer magnetosphere is solved by BATS‐R‐US (Block‐Adaptive Tree Solar wind Roe‐type Upwind Scheme) \cite{gombosi2004}, the inner magnetosphere by the Rice Convection Model (RCM) \cite{toffoletto2003}, and the ionosphere electrodynamics by the Ridley Ionosphere Model (RIM) \cite{ridley2004}. A detailed description of the Geospace model and its modules can be found in \citeA{pulkkinen2013, toth2014}

\subsection{Prediction of $dB/dt$}
\begin{figure}
 \noindent\includegraphics[width=\textwidth]{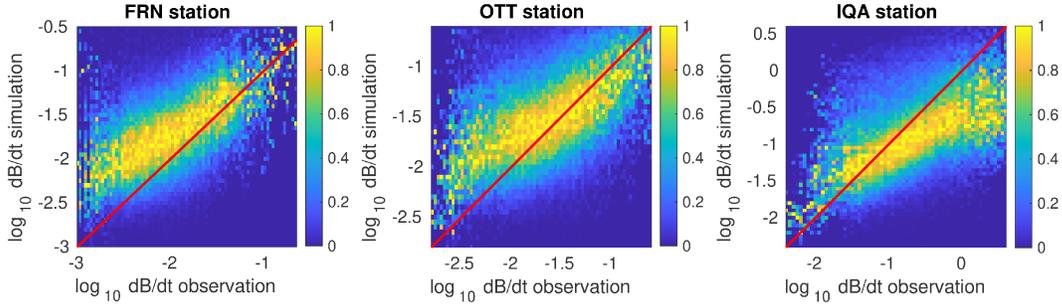}
\caption{2D histogram of the counts of $dB/dt$ as obtained from the Geospace simulation (vertical axis) vs the corresponding measured values (horizontal axis). Both axes are in logarithmic scale, and the heat-map is normalized column-wise with respect to the maximum value for each column, for better visualization. FRN, OTT, and IQA stations are respectively shown in the left, middle and right panels.}\label{fig:hist2D_geo_vs_real_FRN}
\end{figure}

In defining the problem, we follow the strategy introduced in \citeA{pulkkinen2013} and later adopted in \citeA{toth2014} and \citeA{welling2017}.
Specifically, we define
\begin{equation}\label{dbdt_def}
 dB/dt = \textrm{max}_{\{t,t+\Delta t\}} \sqrt{(dB_n/dt)^2+(dB_e/dt)^2}
\end{equation}
as the maximum value of the time derivative of the horizontal magnetic field, over an interval $\Delta t$, where $n$ and $e$ denote the north and east components of the magnetic field, respectively.
More specifically, we restrict the time interval to $\Delta t = 20$ minutes, and we cast the problem as a classification task. Namely, our model predicts the probability that $dB/dt$ will exceed a given threshold at a given location, in a 20-minute interval. {We use overlapping intervals with a 1 minute stride between subsequent intervals}.
Henceforth we simply refer to $dB/dt$ as defined in Eq. (\ref{dbdt_def}).\\
As a proof-of-concept, we will show results for the following three magnetic stations: Fresno, California (Geomagnetic latitude: 43.12$^\circ$N, operated by USGS, code: FRN), Ottawa, Canada (Geomagnetic latitude: 54.88$^\circ$N, operated by GSC, code: OTT), Iqaluit, Canada (Geomagnetic Latitude: 73.25$^\circ$N, operated by GSC, code: IQA), hence testing our new method for low, mid and high magnetic latitudes, respectively. The reported magnetic coordinates are derived from the International Geomagnetic Reference Field (IGRF) 12th generation \cite{thebault2015}.
The extension of this method to any other station is straightforward.\\
The need of enhancing a physics-based approach with machine learning can be appreciated by analyzing the accuracy of the Geospace model in predicting $dB/dt$. Figure \ref{fig:hist2D_geo_vs_real_FRN} shows the number of instances of a given $dB/dt$ value observed in the simulation (vertical axis) versus the corresponding value observed in the data (horizontal axis), both in logarithmic scale (FRN, OTT, and IQA stations shown in the left, middle and right panels, respectively). Each column (i.e. a fixed observed value) is normalized to its maximum value. The statistics are computed over a two year interval (see below).
The solid red line represents a perfect match between predicted and observed values. Figure \ref{fig:hist2D_geo_vs_real_FRN} shows that the simulations tend to underestimate $dB/dt$ for large values (particularly at high latitude) and overestimate it for small values (particularly at low latitude). Also, the range of observed and predicted values is dependent on the geomagnetic latitude, as expected.

The paper is divided as follows. Section 2 introduces the data used for this study and the corresponding time periods covered. Section 3 describes the methodology, including the machine learning technique, the performance metrics, and the features chosen in the model. Section 4 presents the results of the new model, comparing its performance with the output of the Geospace model alone, and emphasizes the probabilistic nature of the forecast. Finally,  in Section 5 we draw conclusions and make final remarks about future directions.  

\section{Data}
The magnetic field historical records have been obtained by the International Real-time Magnetic Observatory Network (INTERMAGNET). The one-minute  data in IAGA-2002 format were retrieved for the period 2001-01-01 to 2019-05-05 (\url{ftp://ftp.seismo.nrcan.gc.ca/intermagnet/minute/variation/IAGA2002/}) for the three stations (FRN, OTT, IQA), consisting of about 9.45M valid entries per station.
The output of the Geospace model used for this work covers the time period 2017-05-28 to 2019-05-05, about 1,000,000 one-minute output values. {In addition, we also use the output of the Geospace model discussed in \citeA{pulkkinen2013}, evaluated over a time period covering 6 geomagnetic storms. Those simulation outputs are made available from the NASA Community Coordinated Modeling Center (CCMC) (\url{https://ccmc.gsfc.nasa.gov/RoR_WWW/pub/dBdt/out/}).} \\	
The Geospace model outputs the magnetic field at the location of the three stations at one minute resolution. However, the inner boundary of the global MHD model is at 2.5 Earth radii ($R_E$). Therefore, the magnetic perturbations at the {geomagnetic observatories} are calculated from the currents using Biot-Savart integrals, taking into account the following three contributions: the currents inside the BATS-R-US domain, the field-aligned currents in the gap region between 1 and 2.5 $R_E$ radial distance, and the Pedersen and Hall currents in the ionosphere electrodynamics model RIM \cite{yu2008}.\\
In order to assess the accuracy of a trained machine learning model, it is important that the performance metrics are calculated on a portion of a data that has not been used for training (so-called unseen data). Moreover, when dealing with temporal dataset, it is equally important that the training and test sets are temporally disjoint so to minimize the temporal correlations between the two and to ensure that the machine learning algorithm does actually learn some patterns and does not merely memorizes the training data. 
{For all our experiments and results, we use as training set the period covering 2017-05-28 to 2018-12-31 and as test set the period 2019-01-01 to 2019-05-05. In addition, three of the six storm events used in \citeA{pulkkinen2013} have been added to the training set (events numbered 1, 3, 4), and the rest have been used for testing. We have verified that the distribution of $dB/dt$ is approximately equal between training and test sets. After excluding gaps in the data, the split results in about 450,000 data points in the training set and 90,000 in the test set.} 

\section{Methodology}
As mentioned in the Introduction, the goal of this work is not to predict the precise value of $dB/dt$ for any given 20 minutes interval, but rather to estimate the probability that a pre-defined threshold will be exceeded. Hence, the first task is to define such thresholds. In this paper, we slightly deviate from \citeA{pulkkinen2013}, which focused on the following four thresholds: $(0.3,0.7,1.1,1.5)$ nT/s, independent of  the station considered. Instead, we define thresholds specific for each location, by analyzing the overall distribution of $dB/dt$ observed in the INTERMAGNET data ($\sim$ 19 years of data) and choosing the following percentiles as thresholds: {60\%, 70\%, 80\%, 90\%, 95\%}. The resulting thresholds are summarized in Table \ref{table:thresholds}.

\begin{table}
 \caption{Thresholds considered for each station (in nT/s)}\label{table:thresholds}
 \centering
 \begin{tabular}{c c c c c c}
 \hline
  Station & 60\% & 70\% & 80\% & 90\% & 95\%  \\
 \hline
 FRN &  0.012  &  0.014  &  0.018 &   0.027 & 0.038\\
 OTT &    0.03 &    0.038 &    0.05 &    0.076 & 0.11\\
 IQA &    0.24 &    0.32 &    0.45 &    0.73& 1.11\\
 \hline
% \multicolumn{2}{l}{$^{a}$Footnote text here.}
 \end{tabular}
 \end{table}

\subsection{Metrics}\label{sec:metrics}
\begin{figure}
 \noindent\includegraphics[width=\textwidth]{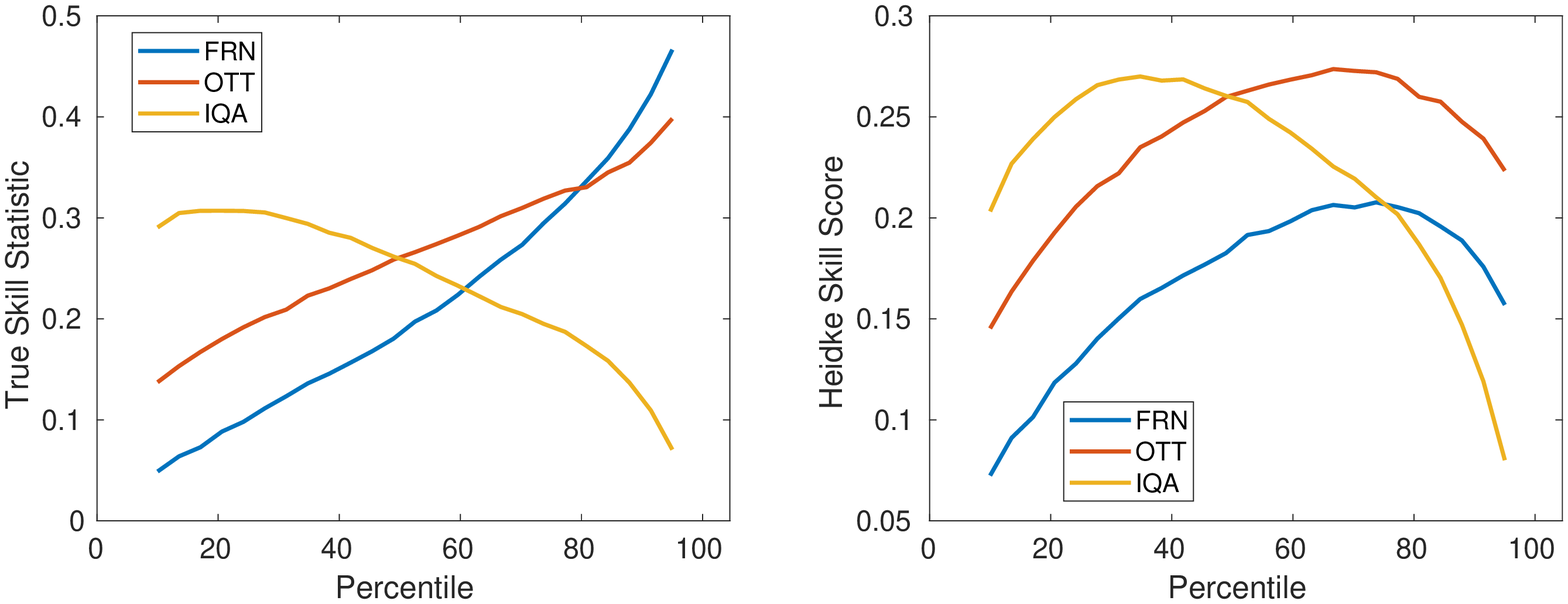}
\caption{True Skill Statistic (left) and Heidke Skill Score (right) obtained from the predictions of the Geospace model, for different stations (in blue for FRN, red for OTT, and yellow for IQA), as functions of the different thresholds percentiles. The percentile are calculated on the distribution of observed $dB/dt$ for a given station over a period of $\sim$ 19 years of data.}
\label{fig:tss_geospace_vs_real}
\end{figure}

The task under consideration is a probabilistic classification: for a given station the model outputs the probability that $dB/dt$ will exceed a specified threshold value. Such a probabilistic outcome can be interpreted as a deterministic binary prediction (i.e. positive/negative) by simply assigning `positive' to all predictions above a certain probability, and `negative' otherwise. Once the probabilistic outcome is interpreted as a binary prediction, one can calculate the following quantities, defined over a certain number of predictions:
\begin{itemize}
 \item $P$ = total number of observed positives (event occurrences);
 \item $N$ = total number of observed negatives (event non-occurrences);
 \item $TP$ = True Positives: number of predicted positives that are observed positives;
 \item $FP$ = False Positives: number of predicted positives that are observed negatives;
 \item $TN$ = True Negatives: number of predicted negatives that are observed negatives;
 \item $FN$ = False Negatives: number of predicted negatives that are observed positives;
\end{itemize}

and the following performance metrics:
\begin{itemize}
 \item $TPR= TP/P$ = True Positive Rate (also called Probability of Detection, Sensitivity, Hit Rate);
 \item $FPR= FP/N$ = False Positive Rate (also called Probability of False Detection, False Alarm Rate);
 \item $TSS = TPR-FPR$ = True Skill Statistics.
 \item $HSS = 2(TP \cdot TN - FN\cdot FP)/(P(FN+TN) + N(TP+FP))$ = Heidke Skill Score
\end{itemize}
The $TPR$ measures the ability to find all positive events and a perfect classifier results in $TPR=1$ ; the $FPR$ measures the probability of wrongly classifying a negative as a positive, and a perfect classifier results in $FPR=0$. Hence, $TSS$ is a useful metric that combines both types of information and should be as close as possible to 1. Moreover, in a Receiver Operating Characteristic (ROC) curve, $TPR$ and $FPR$ are respectively on the vertical and horizontal axis, and $TSS$ measures the distance to the diagonal (no-skill) line \cite{krzanowski2009}.
Finally, the $HSS$ measures the skill of a method compared to a baseline represented by random chance. $HSS$ has been used in \citeA{pulkkinen2013} and is used here for comparison with previous studies.\\
The baseline accuracy is represented by the True Skill Statistic and the Heidke Skill Score yielded by the Geospace model alone, that is by calculating $dB/dt$ directly from the simulation output and comparing to observations. Figure \ref{fig:tss_geospace_vs_real} shows the $TSS$ (left) and $HSS$ (right), where blue, red, and yellow lines are for \emph{FRN}, \emph{OTT}, and \emph{IQA} stations, respectively. {The scores are computed over all the data for which we have Geospace simulation outputs, and are shown for different thresholds, represented on the the horizontal axis in terms of their percentile calculated over $\sim 19$ years of observational data. One can notice that both scores are latitude dependent. Interestingly, while the TSS increases with higher percentiles (less frequent events) for FRN and OTT, the opposite is true for IQA. A different behaviour is also noticible with regards to the HSS score: FRN and OTT stations peak around the 70th percentile, while IQA peaks approximately at the 35th percentile. In general, Figure \ref{fig:tss_geospace_vs_real} shows that the Geospace model performs better at predicting large thresholds of $dB/dt$ at mid and low latitudes than at high latitudes.}

\subsection{Machine Learning classifier}
A variety of methods exist in the Machine Learning arena to perform a probabilistic classification task. For this work, we have opted to use a boosted ensemble of classification trees. The method of choice is called RobustBoost \cite{freund2009}.
In this section we provide a short introduction and appropriate references.\\
Let us assume we want to assign a label $y\in\{0,1\}$ to a data point $\mathbf{x}=\{x_1,x_2,\ldots,x_D\}$, where $D$ is the dimensionality of $\mathbf{x}$. The task is a supervised binary classification, meaning that we make use of a large dataset of labeled examples to infer a pattern between the inputs $\mathbf{x}$ and the binary outputs $y$, that can be used to infer the label of new data points that have not been used to train the model. A decision tree is a simple method that recursively partitions the D-dimensional hyperspace of input variables one dimension at the time, thus creating a tree-like structure. In other words, by taking as decision boundaries hyperplanes defined by simple inequalities such as $x_i<c$, a decision tree divides the input space into a number of hypercubes where a given label is assigned to all the data belonging to the same hypercube. Decision trees have the great advantage of being very transparent and easily interpretable. In fact, one can simply follow the tree structure from top to bottom to understand how a label is associated to a given data point. In order to choose where to set up a decision boundary (i.e. the value of the constant $c$) and along which variable, a partition criterion is followed. Two {standard} partition criteria are the Gini Index, and the Information Gain. The Gini Index measures the reduction in class impurity, which is defined as the probability that two randomly chosen data that belong to the same partition have different labels. At a given iteration when growing a tree, the best partition is the one that reduces such impurity, or in other words that minimizes the probability of a data point being mislabeled. The Information Gain is based on the entropy measured at each node and the optimal split is the one that minimizes the global entropy (or maximizes information). A reference monograph on decision trees is \citeA{breiman2017}.
The accuracy of a classification tree can be improved by using a boosting strategy. Boosting refers to a class of algorithms that makes use of an ensemble of (not very accurate) predictions to produce a much more accurate one. The members of the ensemble are called weak learners and their weighted sum is referred to as strong learner. 
In the context of classification trees, the weak learners are represented by trees that are grown to only a few layers.
One of the most successful and widely applied boosting techniques is \emph{Adaboost} (short for Adaptive Boosting), introduced in the seminal paper by \citeA{freund1997}. This is an algorithm that iteratively adds members to an existing ensemble. The newest member increasingly focuses on the data points that were misclassified by the previous members, and the weights of each member are iteratively adjusted. AdaBoost is typically less prone to overfitting than other algorithms, but it is very sensitive to outliers, because it will keep focusing on the few data points that are mis-classified, eventually at the expense of the remaining dataset. A modification of Adaboost that adds robustness to the algorithm (in the sense of not being so sensitive to outliers) is \emph{RobustBoost}, being introduced in \citeA{freund2009}. RobustBoost can be intuitively understood as ``giving up'' on data points that are so far on the incorrect side of a decision boundary that they are unlikely to be correctly classified even after many iterations.\\
In this work we have used the MATLAB (R2019a) implementation of RobustBoost which is included in the Statistics and Machine Learning Toolbox. We have tested and compared the following boosting techniques: AdaBoost, GentleBoost, LogistBoost, RobustBoost, and Bagging, and although their results were comparable, RobustBoost is the algorithm that consistently yielded better results.

\subsection{Feature selection}\label{sec:features}
\begin{figure}
\hspace*{-2cm}
\noindent\includegraphics[scale=0.5]{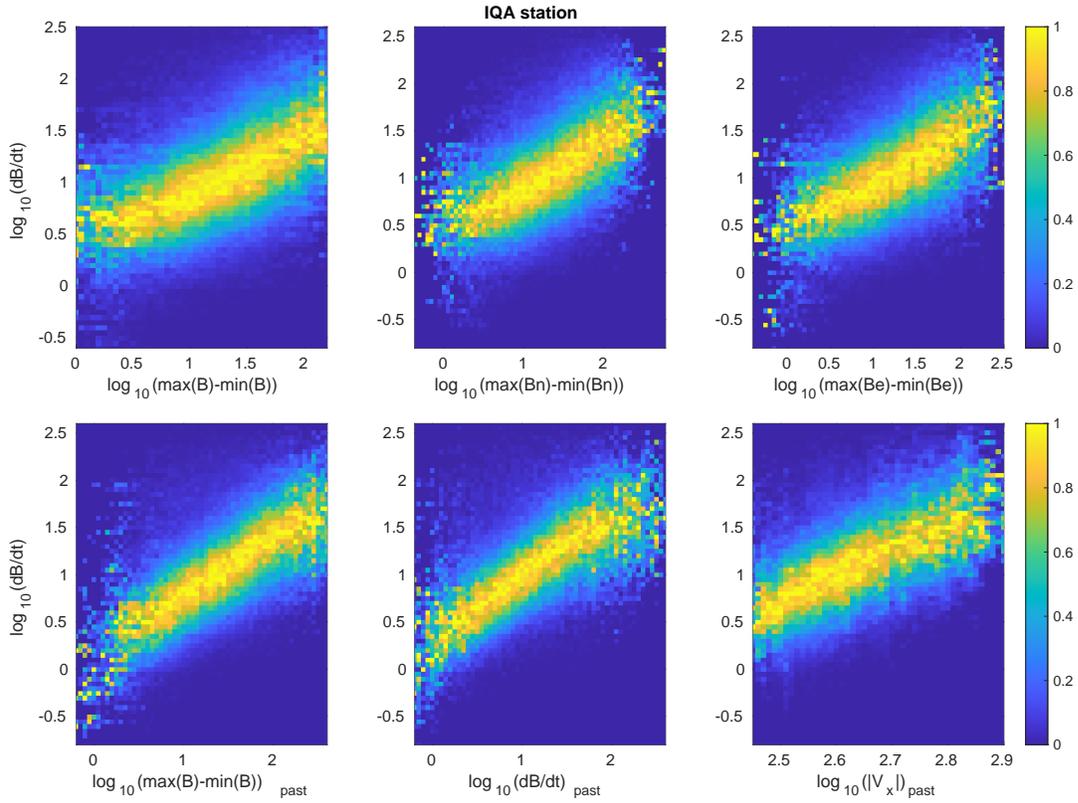}
\caption{2D histogram of the counts of the target variable $dB/dt$ at the IQA station (vertical axis) and the 6 features described in Sec. \ref{sec:features}. Each heat-map is normalized column-wise with respect to its maximum value.}
\label{fig:correlations_IQA}
\end{figure}

\begin{figure}
\centering
\noindent\includegraphics[scale=0.5]{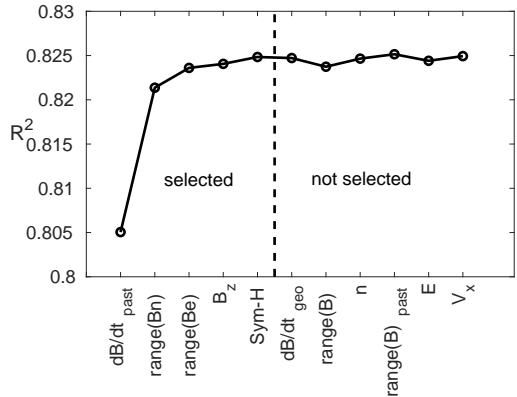}
\caption{Coefficient of determination $R^2$ for the linear models trained succesively on a larger number of features.  Each symbol represents a model trained with the feature reported on horizontal axis in addition to all the features to its left (see Table \ref{tab:features}). The most important features are to the left and the less important to the right.}
\label{fig:feat_selection}
\end{figure}

In the machine learning jargon a feature is an explanatory variable that is used as an input for a given algorithm. 
The present work builds up on the idea presented in \citeA{toth2014}. The main finding was that a strong correlation exists between observed $dB/dt$ and the observed maximum variation in the amplitude of the magnetic field, within the same 20-minute interval. In fact, the correlation is almost linear when both quantities are expressed in their logarithm. \citeA{toth2014} argued that the magnetic perturbations relative to the background dipole value obtained by the Geospace model simulations are much more reliable than the values of $dB/dt$ computed directly from the field.\\
{Feature selection refers to the procedure of selecting the most informative inputs used in a machine learning algorithm, making sure that the number of selected features is large enough for the algorithm to be accurate, but not too large, in order to prevent overfitting and for optimizing computational efficiency. The initial selection of features is done by a visual exploratory analysis of the correlation between candidate features and the target $dB/dt$. We have found that several quantities correlate well when plotted in logarithmic scale. As an example, we show in Figure \ref{fig:correlations_IQA} such correlations for the IQA station. The complete list of initially identified features is in Table \ref{tab:features}. Here, the lead-time of the model forecasts $\Delta T$ are defined as the propagation time of the solar wind between the L1 point (where the solar wind is measured) and the outer boundary of the computational domain, approximately at the Earth's bow shock. In the OMNI dataset used for this study solar wind quantities are conveniently time-shifted to account for the propagation time between L1 and the bow shock. Hence, we have shifted back in time the measurements of Sym-H only, using the timeshift provided by the OMNI dataset (at 1 minute resolution).\\

\begin{sidewaystable}
 \caption{Ranking of features. $T$ denotes the time at which $dB/dt$ is predicted and $\Delta T$ is the solar wind propagation time}\label{tab:features}
 \begin{tabular}{c|c|c|c|c|c}
 \hline
  Rank & Feature & Meaning & Data source & Time & Selected\\
 1 & $\log_{10}(dB/dt)$ & target at previous time& magnetometer  & $T-\Delta T$ & yes \\ 
 2 & $\log_{10}(\max(Bn)-\min(Bn))$ & \multirow{1}{*} Range of North component of magnetic \\ & & field predicted by simulation & Geospace & T & yes \\
 3 & $\log_{10}(\max(Be)-\min(Be))$ & \multirow{1}{*} Range of East component of magnetic \\ & & field predicted by simulation  & Geospace & T & yes \\
 4 & $\log_{10}(Bz)$ &   \multirow{1}{*} z-component (GSM) of \\ & & interplanetary magnetic field & OMNI dataset  & $T-\Delta T$ & yes \\
 5 & $SymH$ & Geomagnetic index Sym-H& OMNI dataset & $T-\Delta T$ & yes \\
 6 & $dB/dt_{geo}$ & target predicted by simulation output & Geospace & T & no \\
 7 &$\log_{10}(\max(B)-\min(B))_{geo}$ &  \multirow{1}{*} Range of magnetic field amplitude \\ & & predicted by simulation  & Geospace & $T-\Delta T$ & no \\
 8 & $\log_{10}(n)$ & solar wind density & OMNI dataset & $T-\Delta T$ & no \\
 9 &$\log_{10}(\max(B)-\min(B))$ & Range of magnetic field amplitude observed& magnetometer & $T-\Delta T$ & no \\  
 10 & $\log_{10}(E)$ & Electric field & OMNI dataset & $T-\Delta T$ & no \\
 11 & $\log_{10}(|V_x|)$ & $x-$ component of solar wind speed & OMNI dataset   & $T-\Delta T$ & no \\
 \hline
 \end{tabular}
\end{sidewaystable}

The procedure chosen to reduce the number of features is known as \emph{backward elimination}, and it works as follows. First, we train a linear regression model using all the features listed in Table \ref{tab:features}, thus assuming the output ($log_{10} (dB/dt)$, measured from magnetometer data) to be a linear combination of the inputs, each weighted by a coefficient. The model returns both the values of the coefficients and their standard deviation. The $t-statistic$ (t-value) is defined as the ratio between coefficients and their standard deviation. Inputs with low t-value (in absolute value) are deemed non-informative. Therefore, we iteratively reduce the number of features by eliminating the one with the smallest t-value and we re-train a new linear model at each iteration with the remaining features. In this way we rank all the features listed in Table \ref{tab:features} (first column). Moreover, for each iteration we record the coefficient of determination $R^2$ as a metric for the goodness of fit. The final ranking of features is represented in Figure \ref{fig:feat_selection}, that shows how the value of $R^2$ changes by increasingly adding features. The features on the horizontal axis are sorted in order of importance from left (most important) to right (less important) and each circle corresponds to a linear model that uses the named feature in addition to all the ones listed to its left. Not surprisingly, the past value of $dB/dt$ is the most informative feature, yielding by itself a $R^2$ value of 0.805. However, the next two are features determined by the Geospace model output, namely the difference between the maximum and minimum values of the North and Easth components of the magnetic field in a 20-minute window. This justifies the grey-box philosophy of combining inputs from simulation outputs with past observations.
On the basis of the backward elimination procedure, we decide to use the top 5 features of Figure \ref{fig:feat_selection}, noticing that $R^2$ tends to plateau with more than 5 features. The procedure has been run on the combined training sets for all three stations. However, to avoid overfitting, at each iteration only 50\% of the combined training set has been used to train the linear model. This explains the small fluctuations of $R^2$ during the plateau, that otherwise would be monotonically increasing, if subsequent models were trained on identical data.
}

\section{Results}
\begin{figure}
 \noindent\includegraphics[width=0.9\textwidth]{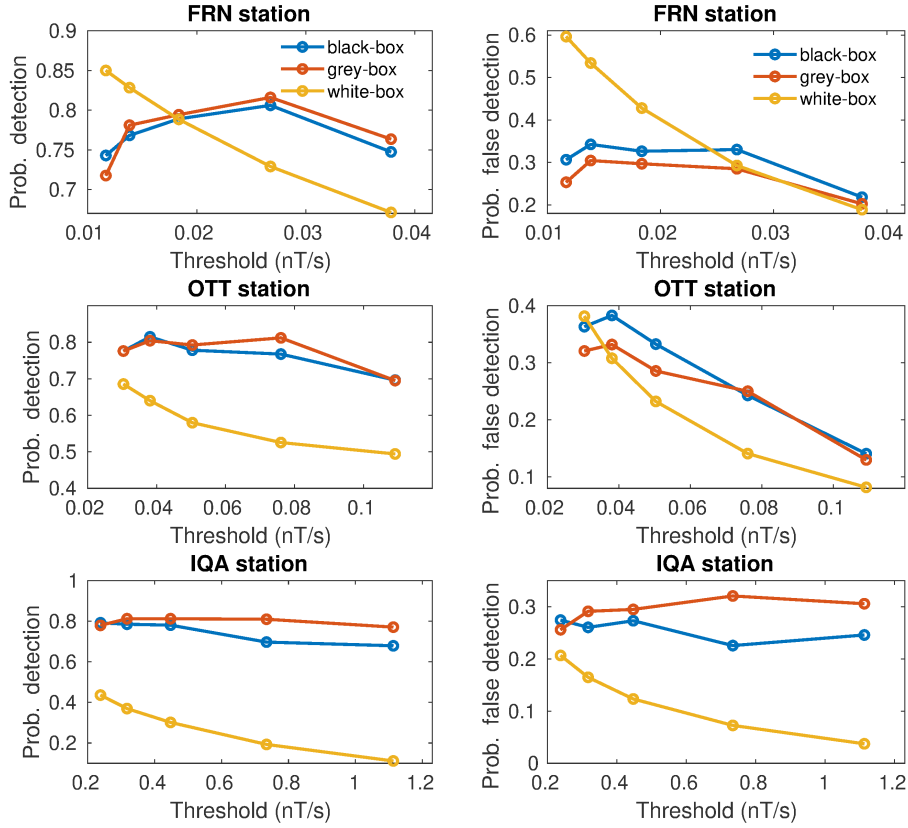}
\caption{Probability of detection (left) and Probability of false detection (right) vs different thresholds (horizontal axis). Blue, red, and yellow lines denote respectively: a black-box model trained without using Geospace output, a gray-box model that uses both past observations and Geospace output, and a white-box model that uses only Geospace outputs. }
\label{fig:pod_stations}
\end{figure}

\begin{figure}
 \noindent\includegraphics[width=0.9\textwidth]{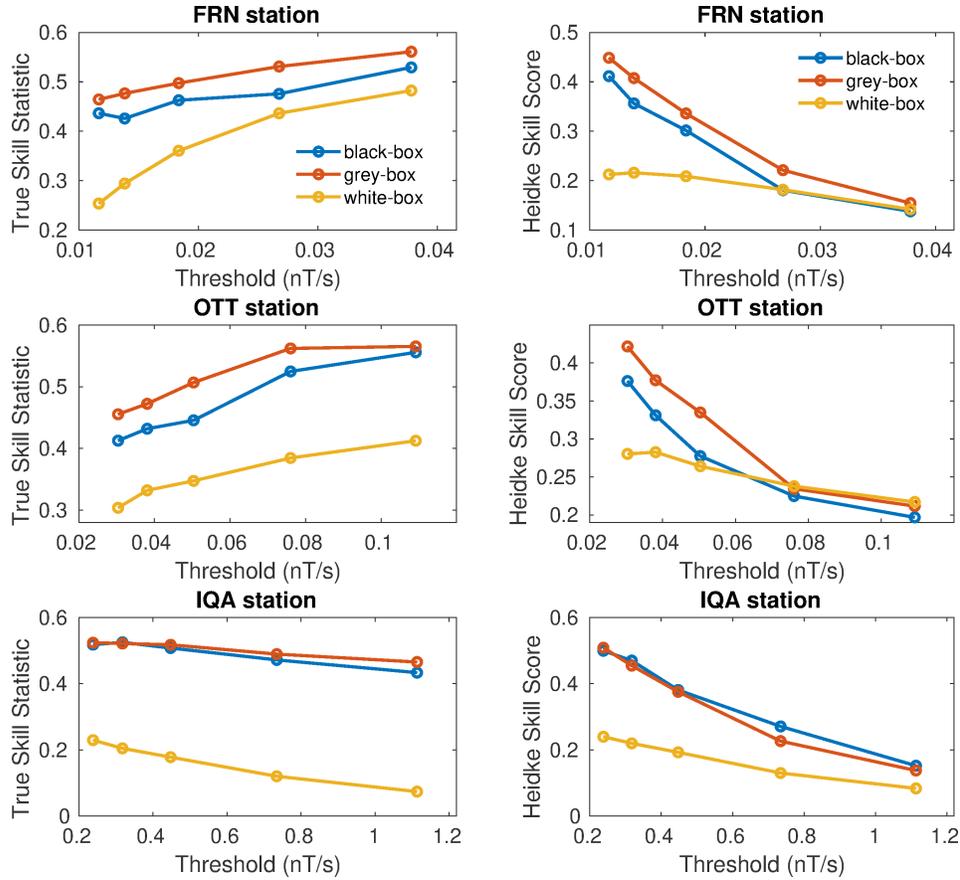}
\caption{True Skill Statistic (left) and Heidke Skill Score (right) vs different threshold (horizontal axis). Blue, red, and yellow lines denote respectively: a black-box model trained without using Geospace output, a gray-box model that uses both past observations and Geospace output, and a white-box model that uses only Geospace outputs. }
\label{fig:tss_stations}
\end{figure}

{In this Section we show the results of our model in terms of the True Positive Rate (TPR, or probability of detection), False Positive Rate (FPR, or probability of false detection), True Skill Statistics, and the Heidke Skill Score discussed in Section \ref{sec:metrics}. Essentially, a different classifier is trained for each station and each threshold. 
The proposed grey-box approach is compared against two alternative approaches: a white-box approach where one simply uses the value of $dB/dt$ predicted by the Geospace model, and a black-box approach where similar machine learning classifiers are trained, with the only difference of not using the inputs coming from the Geospace model. In other words, among the top 5 features listed in Table \ref{tab:features}, the black-box models use only the three that do not come from Geospace output. 
Figure \ref{fig:pod_stations} shows the TPR (left) and FPR (right) for the three stations and as functions of the different threshold levels (see Table \ref{table:thresholds}). 
Figure \ref{fig:tss_stations}  shows TSS (left) and HSS (right) with the same format.  
One can notice that both black- and grey-box models largely outperform the corresponding white-box models. Moreover, although the results are dependent on stations and thresholds, the grey-box model further improves the black-box model, especially for large thresholds, which are the cases of most interest for space weather. On the other hand, whenever the white-box model yields large values for the probability of detection (e.g. for FRN station), the probability of false detection is also large, resulting in low values for both TSS and HSS.
}
\subsection{Re-calibration}\label{sec:recalibration}
\begin{figure}
 \includegraphics[width=1.2\textwidth]{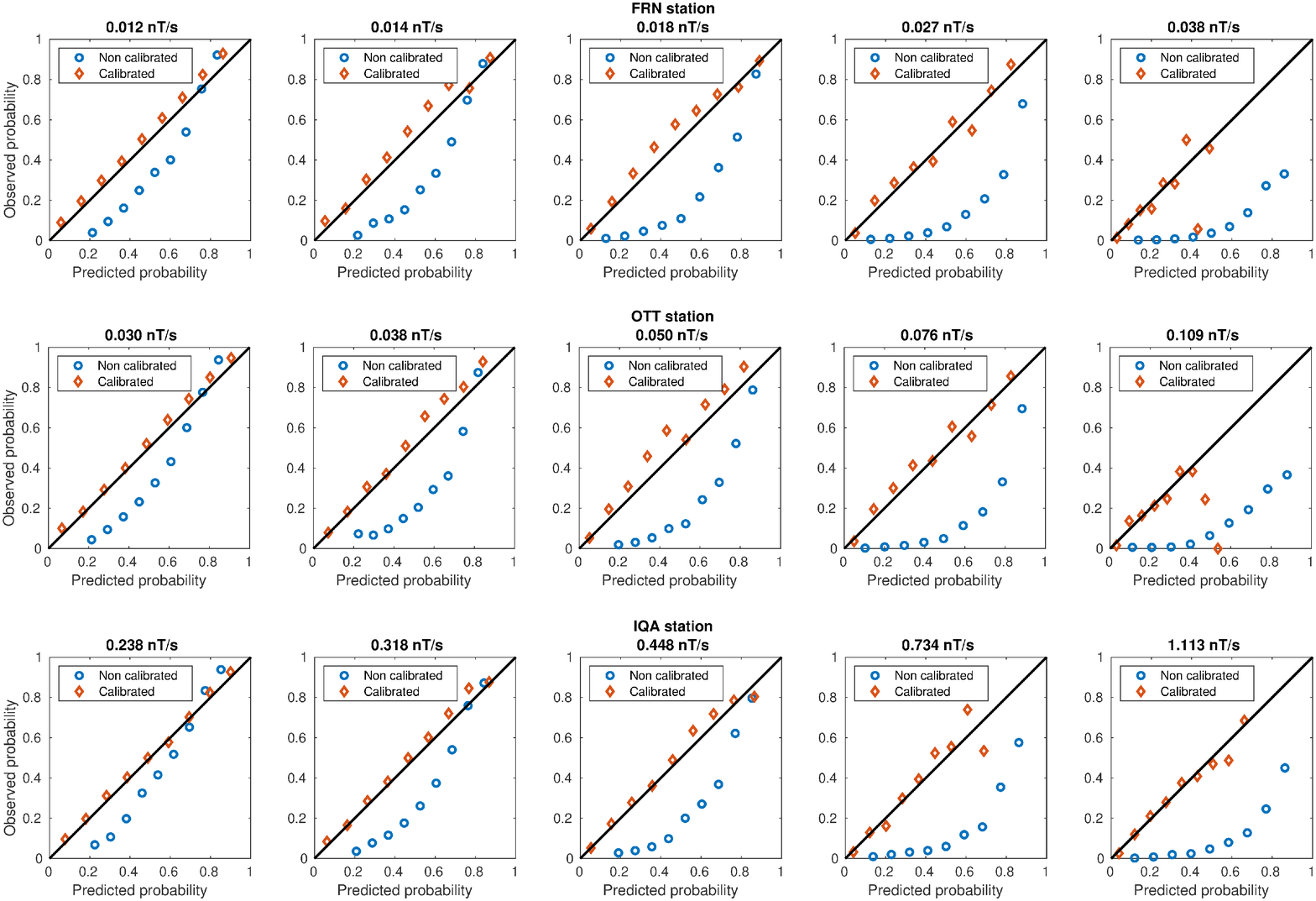}
\caption{Reliability diagrams for different thresholds (increasing from left to right panels). Blue circles indicates the result of the non-calibrated models, and the red diamonds indicate the reliability achieved after re-calibration. FRN, OTT, and IQA stations are shown in the top, middle, bottom row, respectively.}
\label{fig:reliability_FRN}
\end{figure}

As anticipated in the introduction, the goal of this work is not to provide a binary classification, but rather to estimate the probability of exceeding pre-defined thresholds. In principle, classification trees can output probabilities, which are simply calculated as the observed ratio between positives and negatives on a given leaf (the final node on a decision tree) calculated over the whole training set. A well-known problem with classification trees is that such probabilities are often mis-calibrated \cite{niculescu2005}.
Calibration refers to the consistency between the predicted probability assigned to an event and the actual frequency observed for that event. For instance, in the binary classification setting, if we collect all the instances in which a model predicts a probability $p$ for a 'positive' outcome (in our case, exceeding a threshold), that model is well-calibrated if on average a positive is actually observed with frequency $p$ (the frequency being calculated over all those instances). One way to visualize the relationship between predicted probabilities and observed frequency is through a reliability diagram \cite{degroot1983}. To construct such diagram for binary classification, one discretizes the predicted probabilities in bins. For each bin, the average predicted frequency (horizontal axis) is plotted against the true fraction of positive cases in that bin (vertical axis). A perfect calibration will result in a diagonal straight line.
Figure \ref{fig:reliability_FRN} shows the reliability diagrams for FRN, OTT, and IQA, respectively in the top, middle, and bottom row. Each panel refers to a different threshold (see Table \ref{table:thresholds}), and the blue circles represent the calibration of the boosted ensemble models, as trained by the MATLAB routine. One can clearly see that such predictions are mis-calibrated. We apply a simple calibration strategy, where a mapping between old and new probabilities is derived by simply interpolating linearly the blue circles. For instance, a probability of 40\% might be re-calibrated to a new value of 30\%. To perform re-calibration fairly, we have derived the reliability diagram and the corresponding calibration map from the training set only. 
Figure \ref{fig:reliability_FRN} shows the reliability diagram calculated over the test set. The red diamonds represent the re-calibrated reliability diagrams, that clearly suggest that all the models have been properly re-calibrated. The re-calibrated values are the ones that should be used to provide a probabilistic prediction.

\subsection{Receiver operating characteristic (ROC) curve}

\begin{figure}
 \noindent\includegraphics[width=\textwidth]{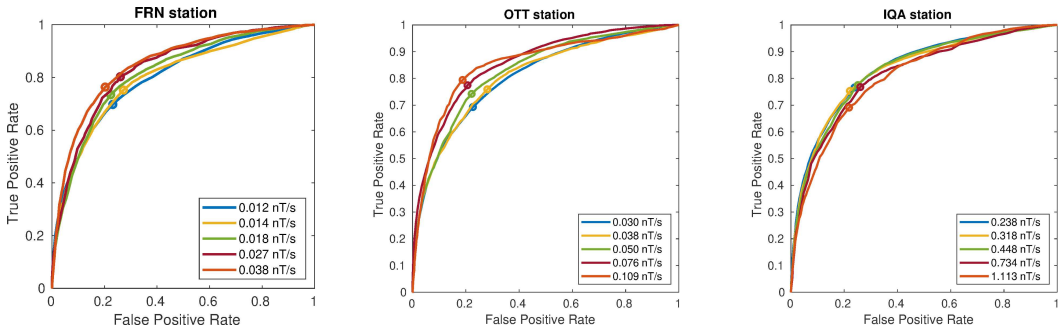}
\caption{ROC curves (TPR vs FPR) for different thresholds. Filled dots indicate the optimal points along a given ROC curve. FRN, OTT, and IQA stations shown in the left, middle, and right panel, respectively.}
\label{fig:ROC_FRN}
\end{figure}

Another important diagnostic for a probabilistic model is the ROC curve. In order to interpret a probabilistic prediction in terms of true/false positives/negatives (see Sec. \ref{sec:metrics}), a probability threshold needs to be used to separate the predicted positives from the negatives. In the limit that such threshold is pushed to 0\%, all the predictions become positives, which means that both the true positive rate (TPR) and the false positive rate (FPR) are equal to 1 (all positives are correctly predicted, but all negative are mis-classified). In the opposite limit, when the threshold is 100\% and all predictions are negative both TPR and FPR become equal to 0 (no positives are predicted, but all negatives are correctly predicted). The ROC curve is a continuous curve in the (FPR,TPR) space that connects these extreme scenarios (TPR=FPR=1 and TPR=FPR=0) by gradually changing the threshold from 0\% to 100\%. The optimal prediction is TPR=1 and FPR=0, and the optimal threshold is the point on the ROC curve with maximum distance from the diagonal.
ROC curves for FRN, OTT, and IQA stations are shown in Figure \ref{fig:ROC_FRN}, respectively in the left, middle, and right panel. Different colors denote the five different thresholds, and a filled circle represents the optimal values (that have been used in previous Figures). Note that the True Skill Statistic (TSS) is the vertical distance between the ROC curve and the diagonal line (TPR=FPR), which represents no skill (i.e. a climatological forecast). The ROC curves demonstrate the general tendency of the models to improve their True Skill Statistic for higher thresholds, as already shown in previous Figures. Moreover, it is important to realize that the re-calibration described in the previous Section does not affect the ROC curve. In fact, by mapping old to new probabilities, the points on a given ROC curve get shifted along the same curve. In other words, what changes through re-calibration is the value of the optimal threshold, but not the corresponding values of TPR, FPR, and Skill Scores. In practice, because the un-calibrated models tend to be overconfident (i.e. below the diagonal line in the reliability diagram), re-calibration changes the optimal threshold from 50\% to larger values. For instance, it can be that for a given model one needs to interpret as positives predictions with probabilities larger than 80\% rather than 50\%.

\section{Conclusions}
\begin{figure}
 \noindent\includegraphics[width=\textwidth]{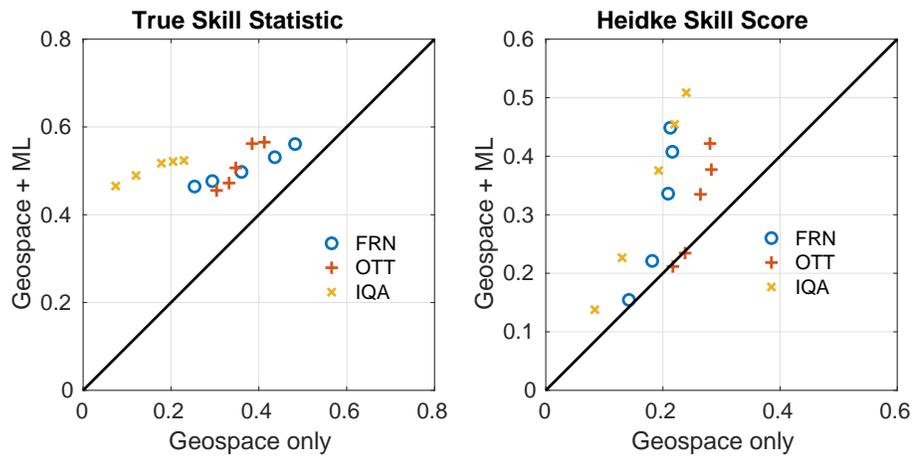}
\caption{Comparison of the True Skill Statistic (left) and Heidke Skill Score (right) for models using the output of the Geospace model alone (horizontal axis) vs the model presented in this paper (combining Geospace outputs with machine learning, vertical axis). The diagonal black line indicates no improvement.}
\label{fig:white_vs_grey_box}
\end{figure}

We have developed  a model that estimates the probability of $dB/dt$ exceeding a given threshold, for three stations ranging from low, to mid and high latitudes (FRN, OTT, and IQA). Five different thresholds were chosen for each station, by calculating the  60, 70, 80, 90, 95 percentiles on a long-span historic dataset ($\sim$ 19 years).
One of the crucial points of this work is that it combines a physics-based prediction provided by the Michigan Geospace model running at SWPC with a machine learning algorithm for binary classification, effectively following what is known as a gray-box approach \cite{camporeale2018,camporeale2019}. Indeed, we have shown that the Geospace model alone provides limited skills for predicting $dB/dt$, although we expect the model to improve over time by better capturing properties of the physical system. 
However, as already noted in \citeA{toth2014}, the maximum perturbation of the magnetic field within a 20-minute interval correlates very strongly with $dB/dt$ and hence it can be used as a predictor in a machine learning algorithm. \\
The chosen machine learning algorithm is an ensemble of classification trees, adaptively boosted via RobustBoost \cite{freund2009}, and the performance metrics that we have analyzed are the True Positive Rate (TPR, or probability of detection), False Positive Rate (FPR, or probability of false detection), True Skill Statistic (TSS), and Heidke Skill Score (HSS). Finally, we have discussed the issue of re-calibration and the ROC curve relative to all models.\\
Overall the gray-box approach proposed in this paper consistently enhances the results of the corresponding white-box approach, where one would directly take the results of the Geospace model as predictors of $dB/dt$. Indeed, Figure \ref{fig:white_vs_grey_box} summarizes the findings of previous Figures by comparing the True Skill Statistic (left panel)  and the Heidke Skill Score (right panel) of the Geospace model alone (horizontal axis) against the corresponding results applying machine learning (vertical axis). Different symbols are for the three different stations, and the region above the diagonal black solid line denotes an improvement.\\

{The new model will be a valuable addition to the operational capabilities of the NOAA's Space Weather Prediction Center. It will be straightforward to extend the model including several stations spanning a range of latitudes and longitudes. We are currently investigating what is the optimal strategy to represent in a compact graphical display the probabilistic outcomes for several stations and several thresholds, such that the SWPC forecasters can extract valuable real-time information on a regional scale and experiment how to incorporate such information in their forecast.}
%%%%%%%%%%%%%%%%%%%%%%%%%%%%%%%%%%%%%%%%%%%%%%%%%%%%%%%%%%%%%%%%
%
%  ACKNOWLEDGMENTS
%
% The acknowledgments must list:
%
% >>>>	A statement that indicates to the reader where the data
% 	supporting the conclusions can be obtained (for example, in the
% 	references, tables, supporting information, and other databases).
%
% 	All funding sources related to this work from all authors
%
% 	Any real or perceived financial conflicts of interests for any
%	author
%
% 	Other affiliations for any author that may be perceived as
% 	having a conflict of interest with respect to the results of this
% 	paper.
%
%
% It is also the appropriate place to thank colleagues and other contributors.
% AGU does not normally allow dedications.

\acknowledgments
The results presented in this paper rely on the data collected at Fresno (FRN), Ottawa (OTT), and Iqaluit (IQA) geomagnetic observatories. We thank the U.S. Geological Survey and Natural Resources Canada Geomagnetism Programs for supporting their operation and INTERMAGNET for promoting high standards of magnetic observatory practice (www.intermagnet.org).  
The INTERMAGNET data used for this study is publicly available on \url{ftp.seismo.nrcan.gc.ca/intermagnet/minute/variation/IAGA2002/}.
The Geospace model outputs discussed in \citeA{pulkkinen2013} are made available by the NASA Community Coordinated Modeling Center (CCMC) (\url{https://ccmc.gsfc.nasa.gov/RoR_WWW/pub/dBdt/out/}).	
\\
All the data and codes will be made available as a Zenodo/Github repository, after the manuscript is accepted for publication.

\clearpage
\newpage

%\end{figure}

\clearpage
\newpage
%\bibliography{camporeale_bib}

\begin{thebibliography}{}

\bibitem [\protect \citeauthoryear {%
Boteler%
\ \BBA {} Pirjola%
}{%
Boteler%
\ \BBA {} Pirjola%
}{%
{\protect \APACyear {1998}}%
}]{%
boteler1998b}
\APACinsertmetastar {%
boteler1998b}%
\begin{APACrefauthors}%
Boteler, D.%
\BCBT {}\ \BBA {} Pirjola, R.%
\end{APACrefauthors}%
\unskip\
\newblock
\APACrefYearMonthDay{1998}{}{}.
\newblock
{\BBOQ}\APACrefatitle {The complex-image method for calculating the magnetic
  and electric fields produced at the surface of the Earth by the auroral
  electrojet} {The complex-image method for calculating the magnetic and
  electric fields produced at the surface of the earth by the auroral
  electrojet}.{\BBCQ}
\newblock
\APACjournalVolNumPages{Geophysical Journal International}{132}{1}{31--40}.
\PrintBackRefs{\CurrentBib}

\bibitem [\protect \citeauthoryear {%
Boteler%
, Pirjola%
\BCBL {}\ \BBA {} Nevanlinna%
}{%
Boteler%
\ \protect \BOthers {.}}{%
{\protect \APACyear {1998}}%
}]{%
boteler1998}
\APACinsertmetastar {%
boteler1998}%
\begin{APACrefauthors}%
Boteler, D.%
, Pirjola, R.%
\BCBL {}\ \BBA {} Nevanlinna, H.%
\end{APACrefauthors}%
\unskip\
\newblock
\APACrefYearMonthDay{1998}{}{}.
\newblock
{\BBOQ}\APACrefatitle {The effects of geomagnetic disturbances on electrical
  systems at the Earth's surface} {The effects of geomagnetic disturbances on
  electrical systems at the earth's surface}.{\BBCQ}
\newblock
\APACjournalVolNumPages{Advances in Space Research}{22}{1}{17--27}.
\PrintBackRefs{\CurrentBib}

\bibitem [\protect \citeauthoryear {%
Breiman%
}{%
Breiman%
}{%
{\protect \APACyear {2017}}%
}]{%
breiman2017}
\APACinsertmetastar {%
breiman2017}%
\begin{APACrefauthors}%
Breiman, L.%
\end{APACrefauthors}%
\unskip\
\newblock
\APACrefYear{2017}.
\newblock
\APACrefbtitle {Classification and regression trees} {Classification and
  regression trees}.
\newblock
\APACaddressPublisher{}{Routledge}.
\PrintBackRefs{\CurrentBib}

\bibitem [\protect \citeauthoryear {%
Camporeale%
}{%
Camporeale%
}{%
{\protect \APACyear {2019}}%
}]{%
camporeale2019}
\APACinsertmetastar {%
camporeale2019}%
\begin{APACrefauthors}%
Camporeale, E.%
\end{APACrefauthors}%
\unskip\
\newblock
\APACrefYearMonthDay{2019}{}{}.
\newblock
{\BBOQ}\APACrefatitle {The Challenge of Machine Learning in Space Weather
  Nowcasting and Forecasting} {The challenge of machine learning in space
  weather nowcasting and forecasting}.{\BBCQ}
\newblock
\APACjournalVolNumPages{Space Weather}{17}{8}{}.
\PrintBackRefs{\CurrentBib}

\bibitem [\protect \citeauthoryear {%
Camporeale%
, Wing%
\BCBL {}\ \BBA {} Johnson%
}{%
Camporeale%
\ \protect \BOthers {.}}{%
{\protect \APACyear {2018}}%
}]{%
camporeale2018}
\APACinsertmetastar {%
camporeale2018}%
\begin{APACrefauthors}%
Camporeale, E.%
, Wing, S.%
\BCBL {}\ \BBA {} Johnson, J.%
\end{APACrefauthors}%
\unskip\
\newblock
\APACrefYear{2018}.
\newblock
\APACrefbtitle {Machine learning techniques for space weather} {Machine
  learning techniques for space weather}.
\newblock
\APACaddressPublisher{}{Elsevier}.
\PrintBackRefs{\CurrentBib}

\bibitem [\protect \citeauthoryear {%
DeGroot%
\ \BBA {} Fienberg%
}{%
DeGroot%
\ \BBA {} Fienberg%
}{%
{\protect \APACyear {1983}}%
}]{%
degroot1983}
\APACinsertmetastar {%
degroot1983}%
\begin{APACrefauthors}%
DeGroot, M\BPBI H.%
\BCBT {}\ \BBA {} Fienberg, S\BPBI E.%
\end{APACrefauthors}%
\unskip\
\newblock
\APACrefYearMonthDay{1983}{}{}.
\newblock
{\BBOQ}\APACrefatitle {The comparison and evaluation of forecasters} {The
  comparison and evaluation of forecasters}.{\BBCQ}
\newblock
\APACjournalVolNumPages{Journal of the Royal Statistical Society: Series D (The
  Statistician)}{32}{1-2}{12--22}.
\PrintBackRefs{\CurrentBib}

\bibitem [\protect \citeauthoryear {%
Freund%
}{%
Freund%
}{%
{\protect \APACyear {2009}}%
}]{%
freund2009}
\APACinsertmetastar {%
freund2009}%
\begin{APACrefauthors}%
Freund, Y.%
\end{APACrefauthors}%
\unskip\
\newblock
\APACrefYearMonthDay{2009}{}{}.
\newblock
{\BBOQ}\APACrefatitle {A more robust boosting algorithm} {A more robust
  boosting algorithm}.{\BBCQ}
\newblock
\APACjournalVolNumPages{arXiv preprint arXiv:0905.2138}{}{}{}.
\PrintBackRefs{\CurrentBib}

\bibitem [\protect \citeauthoryear {%
Freund%
\ \BBA {} Schapire%
}{%
Freund%
\ \BBA {} Schapire%
}{%
{\protect \APACyear {1997}}%
}]{%
freund1997}
\APACinsertmetastar {%
freund1997}%
\begin{APACrefauthors}%
Freund, Y.%
\BCBT {}\ \BBA {} Schapire, R\BPBI E.%
\end{APACrefauthors}%
\unskip\
\newblock
\APACrefYearMonthDay{1997}{}{}.
\newblock
{\BBOQ}\APACrefatitle {A decision-theoretic generalization of on-line learning
  and an application to boosting} {A decision-theoretic generalization of
  on-line learning and an application to boosting}.{\BBCQ}
\newblock
\APACjournalVolNumPages{Journal of computer and system
  sciences}{55}{1}{119--139}.
\PrintBackRefs{\CurrentBib}

\bibitem [\protect \citeauthoryear {%
Gleisner%
\ \BBA {} Lundstedt%
}{%
Gleisner%
\ \BBA {} Lundstedt%
}{%
{\protect \APACyear {2001}}%
}]{%
gleisner2001}
\APACinsertmetastar {%
gleisner2001}%
\begin{APACrefauthors}%
Gleisner, H.%
\BCBT {}\ \BBA {} Lundstedt, H.%
\end{APACrefauthors}%
\unskip\
\newblock
\APACrefYearMonthDay{2001}{}{}.
\newblock
{\BBOQ}\APACrefatitle {A neural network-based local model for prediction of
  geomagnetic disturbances} {A neural network-based local model for prediction
  of geomagnetic disturbances}.{\BBCQ}
\newblock
\APACjournalVolNumPages{Journal of Geophysical Research: Space
  Physics}{106}{A5}{8425--8433}.
\PrintBackRefs{\CurrentBib}

\bibitem [\protect \citeauthoryear {%
Gombosi%
\ \protect \BOthers {.}}{%
Gombosi%
\ \protect \BOthers {.}}{%
{\protect \APACyear {2004}}%
}]{%
gombosi2004}
\APACinsertmetastar {%
gombosi2004}%
\begin{APACrefauthors}%
Gombosi, T\BPBI I.%
, Powell, K\BPBI G.%
, De~Zeeuw, D\BPBI L.%
, Clauer, C\BPBI R.%
, Hansen, K\BPBI C.%
, Manchester, W\BPBI B.%
\BDBL {}others%
\end{APACrefauthors}%
\unskip\
\newblock
\APACrefYearMonthDay{2004}{}{}.
\newblock
{\BBOQ}\APACrefatitle {Solution-adaptive magnetohydrodynamics for space
  plasmas: Sun-to-Earth simulations} {Solution-adaptive magnetohydrodynamics
  for space plasmas: Sun-to-earth simulations}.{\BBCQ}
\newblock
\APACjournalVolNumPages{Computing in science \& engineering}{6}{2}{14}.
\PrintBackRefs{\CurrentBib}

\bibitem [\protect \citeauthoryear {%
Gordeev%
\ \protect \BOthers {.}}{%
Gordeev%
\ \protect \BOthers {.}}{%
{\protect \APACyear {2015}}%
}]{%
gordeev2015}
\APACinsertmetastar {%
gordeev2015}%
\begin{APACrefauthors}%
Gordeev, E.%
, Sergeev, V.%
, Honkonen, I.%
, Kuznetsova, M.%
, Rast{\"a}tter, L.%
, Palmroth, M.%
\BDBL {}Wiltberger, M.%
\end{APACrefauthors}%
\unskip\
\newblock
\APACrefYearMonthDay{2015}{}{}.
\newblock
{\BBOQ}\APACrefatitle {Assessing the performance of community-available global
  MHD models using key system parameters and empirical relationships}
  {Assessing the performance of community-available global mhd models using key
  system parameters and empirical relationships}.{\BBCQ}
\newblock
\APACjournalVolNumPages{Space Weather}{13}{12}{868--884}.
\PrintBackRefs{\CurrentBib}

\bibitem [\protect \citeauthoryear {%
Horton%
, Boteler%
, Overbye%
, Pirjola%
\BCBL {}\ \BBA {} Dugan%
}{%
Horton%
\ \protect \BOthers {.}}{%
{\protect \APACyear {2012}}%
}]{%
horton2012}
\APACinsertmetastar {%
horton2012}%
\begin{APACrefauthors}%
Horton, R.%
, Boteler, D.%
, Overbye, T\BPBI J.%
, Pirjola, R.%
\BCBL {}\ \BBA {} Dugan, R\BPBI C.%
\end{APACrefauthors}%
\unskip\
\newblock
\APACrefYearMonthDay{2012}{}{}.
\newblock
{\BBOQ}\APACrefatitle {A test case for the calculation of geomagnetically
  induced currents} {A test case for the calculation of geomagnetically induced
  currents}.{\BBCQ}
\newblock
\APACjournalVolNumPages{IEEE Transactions on Power
  Delivery}{27}{4}{2368--2373}.
\PrintBackRefs{\CurrentBib}

\bibitem [\protect \citeauthoryear {%
Jordanova%
\ \protect \BOthers {.}}{%
Jordanova%
\ \protect \BOthers {.}}{%
{\protect \APACyear {2018}}%
}]{%
jordanova2018}
\APACinsertmetastar {%
jordanova2018}%
\begin{APACrefauthors}%
Jordanova, V\BPBI K.%
, Delzanno, G\BPBI L.%
, Henderson, M\BPBI G.%
, Godinez, H\BPBI C.%
, Jeffery, C.%
, Lawrence, E\BPBI C.%
\BDBL {}others%
\end{APACrefauthors}%
\unskip\
\newblock
\APACrefYearMonthDay{2018}{}{}.
\newblock
{\BBOQ}\APACrefatitle {Specification of the near-Earth space environment with
  SHIELDS} {Specification of the near-earth space environment with
  shields}.{\BBCQ}
\newblock
\APACjournalVolNumPages{Journal of Atmospheric and Solar-Terrestrial
  Physics}{177}{}{148--159}.
\PrintBackRefs{\CurrentBib}

\bibitem [\protect \citeauthoryear {%
Krzanowski%
\ \BBA {} Hand%
}{%
Krzanowski%
\ \BBA {} Hand%
}{%
{\protect \APACyear {2009}}%
}]{%
krzanowski2009}
\APACinsertmetastar {%
krzanowski2009}%
\begin{APACrefauthors}%
Krzanowski, W\BPBI J.%
\BCBT {}\ \BBA {} Hand, D\BPBI J.%
\end{APACrefauthors}%
\unskip\
\newblock
\APACrefYear{2009}.
\newblock
\APACrefbtitle {ROC curves for continuous data} {Roc curves for continuous
  data}.
\newblock
\APACaddressPublisher{}{Chapman and Hall/CRC}.
\PrintBackRefs{\CurrentBib}

\bibitem [\protect \citeauthoryear {%
Lanzerotti%
}{%
Lanzerotti%
}{%
{\protect \APACyear {2001}}%
}]{%
lanzerotti2001}
\APACinsertmetastar {%
lanzerotti2001}%
\begin{APACrefauthors}%
Lanzerotti, L\BPBI J.%
\end{APACrefauthors}%
\unskip\
\newblock
\APACrefYearMonthDay{2001}{}{}.
\newblock
{\BBOQ}\APACrefatitle {Space weather effects on technologies} {Space weather
  effects on technologies}.{\BBCQ}
\newblock
\APACjournalVolNumPages{Space weather}{125}{}{11--22}.
\PrintBackRefs{\CurrentBib}

\bibitem [\protect \citeauthoryear {%
Lotz%
\ \BBA {} Cilliers%
}{%
Lotz%
\ \BBA {} Cilliers%
}{%
{\protect \APACyear {2015}}%
}]{%
lotz2015}
\APACinsertmetastar {%
lotz2015}%
\begin{APACrefauthors}%
Lotz, S.%
\BCBT {}\ \BBA {} Cilliers, P.%
\end{APACrefauthors}%
\unskip\
\newblock
\APACrefYearMonthDay{2015}{}{}.
\newblock
{\BBOQ}\APACrefatitle {A solar wind-based model of geomagnetic field
  fluctuations at a mid-latitude station} {A solar wind-based model of
  geomagnetic field fluctuations at a mid-latitude station}.{\BBCQ}
\newblock
\APACjournalVolNumPages{Advances in Space Research}{55}{1}{220--230}.
\PrintBackRefs{\CurrentBib}

\bibitem [\protect \citeauthoryear {%
Ngwira%
, Pulkkinen%
, McKinnell%
\BCBL {}\ \BBA {} Cilliers%
}{%
Ngwira%
\ \protect \BOthers {.}}{%
{\protect \APACyear {2008}}%
}]{%
ngwira2008}
\APACinsertmetastar {%
ngwira2008}%
\begin{APACrefauthors}%
Ngwira, C\BPBI M.%
, Pulkkinen, A.%
, McKinnell, L\BHBI A.%
\BCBL {}\ \BBA {} Cilliers, P\BPBI J.%
\end{APACrefauthors}%
\unskip\
\newblock
\APACrefYearMonthDay{2008}{}{}.
\newblock
{\BBOQ}\APACrefatitle {Improved modeling of geomagnetically induced currents in
  the South African power network} {Improved modeling of geomagnetically
  induced currents in the south african power network}.{\BBCQ}
\newblock
\APACjournalVolNumPages{Space Weather}{6}{11}{}.
\PrintBackRefs{\CurrentBib}

\bibitem [\protect \citeauthoryear {%
Niculescu-Mizil%
\ \BBA {} Caruana%
}{%
Niculescu-Mizil%
\ \BBA {} Caruana%
}{%
{\protect \APACyear {2005}}%
}]{%
niculescu2005}
\APACinsertmetastar {%
niculescu2005}%
\begin{APACrefauthors}%
Niculescu-Mizil, A.%
\BCBT {}\ \BBA {} Caruana, R.%
\end{APACrefauthors}%
\unskip\
\newblock
\APACrefYearMonthDay{2005}{}{}.
\newblock
{\BBOQ}\APACrefatitle {Obtaining Calibrated Probabilities from Boosting.}
  {Obtaining calibrated probabilities from boosting.}{\BBCQ}
\newblock
\BIn{} \APACrefbtitle {UAI} {Uai}\ (\BPG~413).
\PrintBackRefs{\CurrentBib}

\bibitem [\protect \citeauthoryear {%
Pirjola%
}{%
Pirjola%
}{%
{\protect \APACyear {2002}}%
}]{%
pirjola2002}
\APACinsertmetastar {%
pirjola2002}%
\begin{APACrefauthors}%
Pirjola, R.%
\end{APACrefauthors}%
\unskip\
\newblock
\APACrefYearMonthDay{2002}{}{}.
\newblock
{\BBOQ}\APACrefatitle {Review on the calculation of surface electric and
  magnetic fields and of geomagnetically induced currents in ground-based
  technological systems} {Review on the calculation of surface electric and
  magnetic fields and of geomagnetically induced currents in ground-based
  technological systems}.{\BBCQ}
\newblock
\APACjournalVolNumPages{Surveys in geophysics}{23}{1}{71--90}.
\PrintBackRefs{\CurrentBib}

\bibitem [\protect \citeauthoryear {%
Pirjola%
}{%
Pirjola%
}{%
{\protect \APACyear {2007}}%
}]{%
pirjola2007}
\APACinsertmetastar {%
pirjola2007}%
\begin{APACrefauthors}%
Pirjola, R.%
\end{APACrefauthors}%
\unskip\
\newblock
\APACrefYearMonthDay{2007}{}{}.
\newblock
{\BBOQ}\APACrefatitle {Space weather effects on power grids} {Space weather
  effects on power grids}.{\BBCQ}
\newblock
\BIn{} \APACrefbtitle {Space Weather-Physics and Effects} {Space
  weather-physics and effects}\ (\BPGS\ 269--288).
\newblock
\APACaddressPublisher{}{Springer}.
\PrintBackRefs{\CurrentBib}

\bibitem [\protect \citeauthoryear {%
Pirjola%
, Boteler%
, Viljanen%
\BCBL {}\ \BBA {} Amm%
}{%
Pirjola%
\ \protect \BOthers {.}}{%
{\protect \APACyear {2000}}%
}]{%
pirjola2000}
\APACinsertmetastar {%
pirjola2000}%
\begin{APACrefauthors}%
Pirjola, R.%
, Boteler, D.%
, Viljanen, A.%
\BCBL {}\ \BBA {} Amm, O.%
\end{APACrefauthors}%
\unskip\
\newblock
\APACrefYearMonthDay{2000}{}{}.
\newblock
{\BBOQ}\APACrefatitle {Prediction of geomagnetically induced currents in power
  transmission systems} {Prediction of geomagnetically induced currents in
  power transmission systems}.{\BBCQ}
\newblock
\APACjournalVolNumPages{Advances in Space Research}{26}{1}{5--14}.
\PrintBackRefs{\CurrentBib}

\bibitem [\protect \citeauthoryear {%
Pulkkinen%
\ \protect \BOthers {.}}{%
Pulkkinen%
\ \protect \BOthers {.}}{%
{\protect \APACyear {2011}}%
}]{%
pulkkinen2011}
\APACinsertmetastar {%
pulkkinen2011}%
\begin{APACrefauthors}%
Pulkkinen, A.%
, Kuznetsova, M.%
, Ridley, A.%
, Raeder, J.%
, Vapirev, A.%
, Weimer, D.%
\BDBL {}others%
\end{APACrefauthors}%
\unskip\
\newblock
\APACrefYearMonthDay{2011}{}{}.
\newblock
{\BBOQ}\APACrefatitle {Geospace environment modeling 2008--2009 challenge:
  Ground magnetic field perturbations} {Geospace environment modeling
  2008--2009 challenge: Ground magnetic field perturbations}.{\BBCQ}
\newblock
\APACjournalVolNumPages{Space Weather}{9}{2}{}.
\PrintBackRefs{\CurrentBib}

\bibitem [\protect \citeauthoryear {%
Pulkkinen%
, Lindahl%
, Viljanen%
\BCBL {}\ \BBA {} Pirjola%
}{%
Pulkkinen%
\ \protect \BOthers {.}}{%
{\protect \APACyear {2005}}%
}]{%
pulkkinen2005}
\APACinsertmetastar {%
pulkkinen2005}%
\begin{APACrefauthors}%
Pulkkinen, A.%
, Lindahl, S.%
, Viljanen, A.%
\BCBL {}\ \BBA {} Pirjola, R.%
\end{APACrefauthors}%
\unskip\
\newblock
\APACrefYearMonthDay{2005}{}{}.
\newblock
{\BBOQ}\APACrefatitle {Geomagnetic storm of 29--31 October 2003:
  Geomagnetically induced currents and their relation to problems in the
  Swedish high-voltage power transmission system} {Geomagnetic storm of 29--31
  october 2003: Geomagnetically induced currents and their relation to problems
  in the swedish high-voltage power transmission system}.{\BBCQ}
\newblock
\APACjournalVolNumPages{Space Weather}{3}{8}{}.
\PrintBackRefs{\CurrentBib}

\bibitem [\protect \citeauthoryear {%
Pulkkinen%
\ \protect \BOthers {.}}{%
Pulkkinen%
\ \protect \BOthers {.}}{%
{\protect \APACyear {2013}}%
}]{%
pulkkinen2013}
\APACinsertmetastar {%
pulkkinen2013}%
\begin{APACrefauthors}%
Pulkkinen, A.%
, Rastatter, L.%
, Kuznetsova, M.%
, Singer, H.%
, Balch, C.%
, Weimer, D.%
\BDBL {}others%
\end{APACrefauthors}%
\unskip\
\newblock
\APACrefYearMonthDay{2013}{}{}.
\newblock
{\BBOQ}\APACrefatitle {Community-wide validation of geospace model ground
  magnetic field perturbation predictions to support model transition to
  operations} {Community-wide validation of geospace model ground magnetic
  field perturbation predictions to support model transition to
  operations}.{\BBCQ}
\newblock
\APACjournalVolNumPages{Space Weather}{11}{6}{369--385}.
\PrintBackRefs{\CurrentBib}

\bibitem [\protect \citeauthoryear {%
Raeder%
, Berchem%
\BCBL {}\ \BBA {} Ashour-Abdalla%
}{%
Raeder%
\ \protect \BOthers {.}}{%
{\protect \APACyear {1998}}%
}]{%
raeder1998}
\APACinsertmetastar {%
raeder1998}%
\begin{APACrefauthors}%
Raeder, J.%
, Berchem, J.%
\BCBL {}\ \BBA {} Ashour-Abdalla, M.%
\end{APACrefauthors}%
\unskip\
\newblock
\APACrefYearMonthDay{1998}{}{}.
\newblock
{\BBOQ}\APACrefatitle {The geospace environment modeling grand challenge:
  Results from a global geospace circulation model} {The geospace environment
  modeling grand challenge: Results from a global geospace circulation
  model}.{\BBCQ}
\newblock
\APACjournalVolNumPages{Journal of Geophysical Research: Space
  Physics}{103}{A7}{14787--14797}.
\PrintBackRefs{\CurrentBib}

\bibitem [\protect \citeauthoryear {%
Rast{\"a}tter%
\ \protect \BOthers {.}}{%
Rast{\"a}tter%
\ \protect \BOthers {.}}{%
{\protect \APACyear {2013}}%
}]{%
rastatter2013}
\APACinsertmetastar {%
rastatter2013}%
\begin{APACrefauthors}%
Rast{\"a}tter, L.%
, Kuznetsova, M.%
, Glocer, A.%
, Welling, D.%
, Meng, X.%
, Raeder, J.%
\BDBL {}others%
\end{APACrefauthors}%
\unskip\
\newblock
\APACrefYearMonthDay{2013}{}{}.
\newblock
{\BBOQ}\APACrefatitle {Geospace environment modeling 2008--2009 challenge: D st
  index} {Geospace environment modeling 2008--2009 challenge: D st
  index}.{\BBCQ}
\newblock
\APACjournalVolNumPages{Space Weather}{11}{4}{187--205}.
\PrintBackRefs{\CurrentBib}

\bibitem [\protect \citeauthoryear {%
Rast{\"a}tter%
\ \protect \BOthers {.}}{%
Rast{\"a}tter%
\ \protect \BOthers {.}}{%
{\protect \APACyear {2011}}%
}]{%
rastatter2011}
\APACinsertmetastar {%
rastatter2011}%
\begin{APACrefauthors}%
Rast{\"a}tter, L.%
, Kuznetsova, M.%
, Vapirev, A.%
, Ridley, A.%
, Wiltberger, M.%
, Pulkkinen, A.%
\BDBL {}Singer, H.%
\end{APACrefauthors}%
\unskip\
\newblock
\APACrefYearMonthDay{2011}{}{}.
\newblock
{\BBOQ}\APACrefatitle {Geospace environment modeling 2008--2009 challenge:
  Geosynchronous magnetic field} {Geospace environment modeling 2008--2009
  challenge: Geosynchronous magnetic field}.{\BBCQ}
\newblock
\APACjournalVolNumPages{Space Weather}{9}{4}{1--15}.
\PrintBackRefs{\CurrentBib}

\bibitem [\protect \citeauthoryear {%
Ridley%
, Gombosi%
\BCBL {}\ \BBA {} DeZeeuw%
}{%
Ridley%
\ \protect \BOthers {.}}{%
{\protect \APACyear {2004}}%
}]{%
ridley2004}
\APACinsertmetastar {%
ridley2004}%
\begin{APACrefauthors}%
Ridley, A.%
, Gombosi, T.%
\BCBL {}\ \BBA {} DeZeeuw, D.%
\end{APACrefauthors}%
\unskip\
\newblock
\APACrefYearMonthDay{2004}{}{}.
\newblock
{\BBOQ}\APACrefatitle {Ionospheric control of the magnetosphere: Conductance}
  {Ionospheric control of the magnetosphere: Conductance}.{\BBCQ}
\newblock
\BIn{} \APACrefbtitle {Annales Geophysicae} {Annales geophysicae}\ (\BVOL~22,
  \BPGS\ 567--584).
\PrintBackRefs{\CurrentBib}

\bibitem [\protect \citeauthoryear {%
Schrijver%
\ \BBA {} Mitchell%
}{%
Schrijver%
\ \BBA {} Mitchell%
}{%
{\protect \APACyear {2013}}%
}]{%
schrijver2013}
\APACinsertmetastar {%
schrijver2013}%
\begin{APACrefauthors}%
Schrijver, C\BPBI J.%
\BCBT {}\ \BBA {} Mitchell, S\BPBI D.%
\end{APACrefauthors}%
\unskip\
\newblock
\APACrefYearMonthDay{2013}{}{}.
\newblock
{\BBOQ}\APACrefatitle {Disturbances in the US electric grid associated with
  geomagnetic activity} {Disturbances in the us electric grid associated with
  geomagnetic activity}.{\BBCQ}
\newblock
\APACjournalVolNumPages{Journal of Space Weather and Space Climate}{3}{}{A19}.
\PrintBackRefs{\CurrentBib}

\bibitem [\protect \citeauthoryear {%
Th{\'e}bault%
\ \protect \BOthers {.}}{%
Th{\'e}bault%
\ \protect \BOthers {.}}{%
{\protect \APACyear {2015}}%
}]{%
thebault2015}
\APACinsertmetastar {%
thebault2015}%
\begin{APACrefauthors}%
Th{\'e}bault, E.%
, Finlay, C\BPBI C.%
, Beggan, C\BPBI D.%
, Alken, P.%
, Aubert, J.%
, Barrois, O.%
\BDBL {}others%
\end{APACrefauthors}%
\unskip\
\newblock
\APACrefYearMonthDay{2015}{}{}.
\newblock
{\BBOQ}\APACrefatitle {International geomagnetic reference field: the 12th
  generation} {International geomagnetic reference field: the 12th
  generation}.{\BBCQ}
\newblock
\APACjournalVolNumPages{Earth, Planets and Space}{67}{1}{79}.
\PrintBackRefs{\CurrentBib}

\bibitem [\protect \citeauthoryear {%
Toffoletto%
, Sazykin%
, Spiro%
\BCBL {}\ \BBA {} Wolf%
}{%
Toffoletto%
\ \protect \BOthers {.}}{%
{\protect \APACyear {2003}}%
}]{%
toffoletto2003}
\APACinsertmetastar {%
toffoletto2003}%
\begin{APACrefauthors}%
Toffoletto, F.%
, Sazykin, S.%
, Spiro, R.%
\BCBL {}\ \BBA {} Wolf, R.%
\end{APACrefauthors}%
\unskip\
\newblock
\APACrefYearMonthDay{2003}{}{}.
\newblock
{\BBOQ}\APACrefatitle {Inner magnetospheric modeling with the Rice Convection
  Model} {Inner magnetospheric modeling with the rice convection model}.{\BBCQ}
\newblock
\APACjournalVolNumPages{Space Science Reviews}{107}{1-2}{175--196}.
\PrintBackRefs{\CurrentBib}

\bibitem [\protect \citeauthoryear {%
T{\'o}th%
, Meng%
, Gombosi%
\BCBL {}\ \BBA {} Rast{\"a}tter%
}{%
T{\'o}th%
\ \protect \BOthers {.}}{%
{\protect \APACyear {2014}}%
}]{%
toth2014}
\APACinsertmetastar {%
toth2014}%
\begin{APACrefauthors}%
T{\'o}th, G.%
, Meng, X.%
, Gombosi, T\BPBI I.%
\BCBL {}\ \BBA {} Rast{\"a}tter, L.%
\end{APACrefauthors}%
\unskip\
\newblock
\APACrefYearMonthDay{2014}{}{}.
\newblock
{\BBOQ}\APACrefatitle {Predicting the time derivative of local magnetic
  perturbations} {Predicting the time derivative of local magnetic
  perturbations}.{\BBCQ}
\newblock
\APACjournalVolNumPages{Journal of Geophysical Research: Space
  Physics}{119}{1}{310--321}.
\PrintBackRefs{\CurrentBib}

\bibitem [\protect \citeauthoryear {%
T{\'o}th%
\ \protect \BOthers {.}}{%
T{\'o}th%
\ \protect \BOthers {.}}{%
{\protect \APACyear {2005}}%
}]{%
toth2005}
\APACinsertmetastar {%
toth2005}%
\begin{APACrefauthors}%
T{\'o}th, G.%
, Sokolov, I\BPBI V.%
, Gombosi, T\BPBI I.%
, Chesney, D\BPBI R.%
, Clauer, C\BPBI R.%
, De~Zeeuw, D\BPBI L.%
\BDBL {}others%
\end{APACrefauthors}%
\unskip\
\newblock
\APACrefYearMonthDay{2005}{}{}.
\newblock
{\BBOQ}\APACrefatitle {Space Weather Modeling Framework: A new tool for the
  space science community} {Space weather modeling framework: A new tool for
  the space science community}.{\BBCQ}
\newblock
\APACjournalVolNumPages{Journal of Geophysical Research: Space
  Physics}{110}{A12}{}.
\PrintBackRefs{\CurrentBib}

\bibitem [\protect \citeauthoryear {%
T{\'o}th%
\ \protect \BOthers {.}}{%
T{\'o}th%
\ \protect \BOthers {.}}{%
{\protect \APACyear {2012}}%
}]{%
toth2012}
\APACinsertmetastar {%
toth2012}%
\begin{APACrefauthors}%
T{\'o}th, G.%
, Van~der Holst, B.%
, Sokolov, I\BPBI V.%
, De~Zeeuw, D\BPBI L.%
, Gombosi, T\BPBI I.%
, Fang, F.%
\BDBL {}others%
\end{APACrefauthors}%
\unskip\
\newblock
\APACrefYearMonthDay{2012}{}{}.
\newblock
{\BBOQ}\APACrefatitle {Adaptive numerical algorithms in space weather modeling}
  {Adaptive numerical algorithms in space weather modeling}.{\BBCQ}
\newblock
\APACjournalVolNumPages{Journal of Computational Physics}{231}{3}{870--903}.
\PrintBackRefs{\CurrentBib}

\bibitem [\protect \citeauthoryear {%
Viljanen%
}{%
Viljanen%
}{%
{\protect \APACyear {1997}}%
}]{%
viljanen1997}
\APACinsertmetastar {%
viljanen1997}%
\begin{APACrefauthors}%
Viljanen, A.%
\end{APACrefauthors}%
\unskip\
\newblock
\APACrefYearMonthDay{1997}{}{}.
\newblock
{\BBOQ}\APACrefatitle {The relation between geomagnetic variations and their
  time derivatives and implications for estimation of induction risks} {The
  relation between geomagnetic variations and their time derivatives and
  implications for estimation of induction risks}.{\BBCQ}
\newblock
\APACjournalVolNumPages{Geophysical research letters}{24}{6}{631--634}.
\PrintBackRefs{\CurrentBib}

\bibitem [\protect \citeauthoryear {%
Viljanen%
, Nevanlinna%
, Pajunp{\"a}{\"a}%
\BCBL {}\ \BBA {} Pulkkinen%
}{%
Viljanen%
\ \protect \BOthers {.}}{%
{\protect \APACyear {2001}}%
}]{%
viljanen2001}
\APACinsertmetastar {%
viljanen2001}%
\begin{APACrefauthors}%
Viljanen, A.%
, Nevanlinna, H.%
, Pajunp{\"a}{\"a}, K.%
\BCBL {}\ \BBA {} Pulkkinen, A.%
\end{APACrefauthors}%
\unskip\
\newblock
\APACrefYearMonthDay{2001}{}{}.
\newblock
{\BBOQ}\APACrefatitle {Time derivative of the horizontal geomagnetic field as
  an activity indicator} {Time derivative of the horizontal geomagnetic field
  as an activity indicator}.{\BBCQ}
\newblock
\BIn{} \APACrefbtitle {Annales Geophysicae} {Annales geophysicae}\ (\BVOL~19,
  \BPGS\ 1107--1118).
\PrintBackRefs{\CurrentBib}

\bibitem [\protect \citeauthoryear {%
Viljanen%
, Pulkkinen%
, Amm%
, Pirjola%
\BCBL {}\ \BBA {} Korja%
}{%
Viljanen%
\ \protect \BOthers {.}}{%
{\protect \APACyear {2004}}%
}]{%
viljanen2004}
\APACinsertmetastar {%
viljanen2004}%
\begin{APACrefauthors}%
Viljanen, A.%
, Pulkkinen, A.%
, Amm, O.%
, Pirjola, R.%
\BCBL {}\ \BBA {} Korja, T.%
\end{APACrefauthors}%
\unskip\
\newblock
\APACrefYearMonthDay{2004}{}{}.
\newblock
{\BBOQ}\APACrefatitle {Fast computation of the geoelectric field using the
  method of elementary current systems and planar Earth models} {Fast
  computation of the geoelectric field using the method of elementary current
  systems and planar earth models}.{\BBCQ}
\newblock
\BIn{} \APACrefbtitle {Annales Geophysicae} {Annales geophysicae}\ (\BVOL~22,
  \BPGS\ 101--113).
\PrintBackRefs{\CurrentBib}

\bibitem [\protect \citeauthoryear {%
Weigel%
, Klimas%
\BCBL {}\ \BBA {} Vassiliadis%
}{%
Weigel%
\ \protect \BOthers {.}}{%
{\protect \APACyear {2003}}%
}]{%
weigel2003}
\APACinsertmetastar {%
weigel2003}%
\begin{APACrefauthors}%
Weigel, R.%
, Klimas, A.%
\BCBL {}\ \BBA {} Vassiliadis, D.%
\end{APACrefauthors}%
\unskip\
\newblock
\APACrefYearMonthDay{2003}{}{}.
\newblock
{\BBOQ}\APACrefatitle {Solar wind coupling to and predictability of ground
  magnetic fields and their time derivatives} {Solar wind coupling to and
  predictability of ground magnetic fields and their time derivatives}.{\BBCQ}
\newblock
\APACjournalVolNumPages{Journal of Geophysical Research: Space
  Physics}{108}{A7}{}.
\PrintBackRefs{\CurrentBib}

\bibitem [\protect \citeauthoryear {%
Weigel%
, Vassiliadis%
\BCBL {}\ \BBA {} Klimas%
}{%
Weigel%
\ \protect \BOthers {.}}{%
{\protect \APACyear {2002}}%
}]{%
weigel2002}
\APACinsertmetastar {%
weigel2002}%
\begin{APACrefauthors}%
Weigel, R.%
, Vassiliadis, D.%
\BCBL {}\ \BBA {} Klimas, A.%
\end{APACrefauthors}%
\unskip\
\newblock
\APACrefYearMonthDay{2002}{}{}.
\newblock
{\BBOQ}\APACrefatitle {Coupling of the solar wind to temporal fluctuations in
  ground magnetic fields} {Coupling of the solar wind to temporal fluctuations
  in ground magnetic fields}.{\BBCQ}
\newblock
\APACjournalVolNumPages{Geophysical Research Letters}{29}{19}{21--1}.
\PrintBackRefs{\CurrentBib}

\bibitem [\protect \citeauthoryear {%
Weimer%
}{%
Weimer%
}{%
{\protect \APACyear {2013}}%
}]{%
weimer2013}
\APACinsertmetastar {%
weimer2013}%
\begin{APACrefauthors}%
Weimer, D\BPBI R.%
\end{APACrefauthors}%
\unskip\
\newblock
\APACrefYearMonthDay{2013}{}{}.
\newblock
{\BBOQ}\APACrefatitle {An empirical model of ground-level geomagnetic
  perturbations} {An empirical model of ground-level geomagnetic
  perturbations}.{\BBCQ}
\newblock
\APACjournalVolNumPages{Space Weather}{11}{3}{107--120}.
\PrintBackRefs{\CurrentBib}

\bibitem [\protect \citeauthoryear {%
Welling%
}{%
Welling%
}{%
{\protect \APACyear {2019}}%
}]{%
welling2019}
\APACinsertmetastar {%
welling2019}%
\begin{APACrefauthors}%
Welling, D.%
\end{APACrefauthors}%
\unskip\
\newblock
\APACrefYearMonthDay{2019}{}{}.
\newblock
{\BBOQ}\APACrefatitle {Magnetohydrodynamic Models of B and Their Use in GIC
  Estimates} {Magnetohydrodynamic models of b and their use in gic
  estimates}.{\BBCQ}
\newblock
\APACjournalVolNumPages{Geomagnetically Induced Currents from the Sun to the
  Power Grid}{}{}{43--65}.
\PrintBackRefs{\CurrentBib}

\bibitem [\protect \citeauthoryear {%
Welling%
, Anderson%
, Crowley%
, Pulkkinen%
\BCBL {}\ \BBA {} Rast{\"a}tter%
}{%
Welling%
\ \protect \BOthers {.}}{%
{\protect \APACyear {2017}}%
}]{%
welling2017}
\APACinsertmetastar {%
welling2017}%
\begin{APACrefauthors}%
Welling, D.%
, Anderson, B.%
, Crowley, G.%
, Pulkkinen, A.%
\BCBL {}\ \BBA {} Rast{\"a}tter, L.%
\end{APACrefauthors}%
\unskip\
\newblock
\APACrefYearMonthDay{2017}{}{}.
\newblock
{\BBOQ}\APACrefatitle {Exploring predictive performance: A reanalysis of the
  geospace model transition challenge} {Exploring predictive performance: A
  reanalysis of the geospace model transition challenge}.{\BBCQ}
\newblock
\APACjournalVolNumPages{Space Weather}{15}{1}{192--203}.
\PrintBackRefs{\CurrentBib}

\bibitem [\protect \citeauthoryear {%
Welling%
\ \BBA {} Ridley%
}{%
Welling%
\ \BBA {} Ridley%
}{%
{\protect \APACyear {2010}}%
}]{%
welling2010}
\APACinsertmetastar {%
welling2010}%
\begin{APACrefauthors}%
Welling, D.%
\BCBT {}\ \BBA {} Ridley, A.%
\end{APACrefauthors}%
\unskip\
\newblock
\APACrefYearMonthDay{2010}{}{}.
\newblock
{\BBOQ}\APACrefatitle {Validation of SWMF magnetic field and plasma}
  {Validation of swmf magnetic field and plasma}.{\BBCQ}
\newblock
\APACjournalVolNumPages{Space Weather}{8}{3}{}.
\PrintBackRefs{\CurrentBib}

\bibitem [\protect \citeauthoryear {%
Wintoft%
}{%
Wintoft%
}{%
{\protect \APACyear {2005}}%
}]{%
wintoft2005}
\APACinsertmetastar {%
wintoft2005}%
\begin{APACrefauthors}%
Wintoft, P.%
\end{APACrefauthors}%
\unskip\
\newblock
\APACrefYearMonthDay{2005}{}{}.
\newblock
{\BBOQ}\APACrefatitle {Study of the solar wind coupling to the time difference
  horizontal geomagnetic field} {Study of the solar wind coupling to the time
  difference horizontal geomagnetic field}.{\BBCQ}
\newblock
\BIn{} \APACrefbtitle {Annales Geophysicae} {Annales geophysicae}\ (\BVOL~23,
  \BPGS\ 1949--1957).
\PrintBackRefs{\CurrentBib}

\bibitem [\protect \citeauthoryear {%
Wintoft%
, Wik%
, Lundstedt%
\BCBL {}\ \BBA {} Eliasson%
}{%
Wintoft%
\ \protect \BOthers {.}}{%
{\protect \APACyear {2005}}%
}]{%
wintoft2005b}
\APACinsertmetastar {%
wintoft2005b}%
\begin{APACrefauthors}%
Wintoft, P.%
, Wik, M.%
, Lundstedt, H.%
\BCBL {}\ \BBA {} Eliasson, L.%
\end{APACrefauthors}%
\unskip\
\newblock
\APACrefYearMonthDay{2005}{}{}.
\newblock
{\BBOQ}\APACrefatitle {Predictions of local ground geomagnetic field
  fluctuations during the 7--10 November 2004 events studied with solar wind
  driven models} {Predictions of local ground geomagnetic field fluctuations
  during the 7--10 november 2004 events studied with solar wind driven
  models}.{\BBCQ}
\newblock
\BIn{} \APACrefbtitle {Annales Geophysicae} {Annales geophysicae}\ (\BVOL~23,
  \BPGS\ 3095--3101).
\PrintBackRefs{\CurrentBib}

\bibitem [\protect \citeauthoryear {%
Wintoft%
, Wik%
\BCBL {}\ \BBA {} Viljanen%
}{%
Wintoft%
\ \protect \BOthers {.}}{%
{\protect \APACyear {2015}}%
}]{%
wintoft2015}
\APACinsertmetastar {%
wintoft2015}%
\begin{APACrefauthors}%
Wintoft, P.%
, Wik, M.%
\BCBL {}\ \BBA {} Viljanen, A.%
\end{APACrefauthors}%
\unskip\
\newblock
\APACrefYearMonthDay{2015}{}{}.
\newblock
{\BBOQ}\APACrefatitle {Solar wind driven empirical forecast models of the time
  derivative of the ground magnetic field} {Solar wind driven empirical
  forecast models of the time derivative of the ground magnetic field}.{\BBCQ}
\newblock
\APACjournalVolNumPages{Journal of Space Weather and Space Climate}{5}{}{A7}.
\PrintBackRefs{\CurrentBib}

\bibitem [\protect \citeauthoryear {%
Yu%
\ \BBA {} Ridley%
}{%
Yu%
\ \BBA {} Ridley%
}{%
{\protect \APACyear {2008}}%
}]{%
yu2008}
\APACinsertmetastar {%
yu2008}%
\begin{APACrefauthors}%
Yu, Y.%
\BCBT {}\ \BBA {} Ridley, A\BPBI J.%
\end{APACrefauthors}%
\unskip\
\newblock
\APACrefYearMonthDay{2008}{}{}.
\newblock
{\BBOQ}\APACrefatitle {Validation of the space weather modeling framework using
  ground-based magnetometers} {Validation of the space weather modeling
  framework using ground-based magnetometers}.{\BBCQ}
\newblock
\APACjournalVolNumPages{Space Weather}{6}{5}{1--20}.
\PrintBackRefs{\CurrentBib}

\bibitem [\protect \citeauthoryear {%
Zhang%
\ \protect \BOthers {.}}{%
Zhang%
\ \protect \BOthers {.}}{%
{\protect \APACyear {2019}}%
}]{%
zhang2019}
\APACinsertmetastar {%
zhang2019}%
\begin{APACrefauthors}%
Zhang, B.%
, Sorathia, K\BPBI A.%
, Lyon, J\BPBI G.%
, Merkin, V\BPBI G.%
, Garretson, J\BPBI S.%
\BCBL {}\ \BBA {} Wiltberger, M.%
\end{APACrefauthors}%
\unskip\
\newblock
\APACrefYearMonthDay{2019}{}{}.
\newblock
{\BBOQ}\APACrefatitle {GAMERA: A Three-dimensional Finite-volume MHD Solver for
  Non-orthogonal Curvilinear Geometries} {Gamera: A three-dimensional
  finite-volume mhd solver for non-orthogonal curvilinear geometries}.{\BBCQ}
\newblock
\APACjournalVolNumPages{The Astrophysical Journal Supplement
  Series}{244}{1}{20}.
\PrintBackRefs{\CurrentBib}

\end{thebibliography}

%Reference citation instructions and examples:
%
% Please use ONLY \cite and \citeA for reference citations.
% \cite for parenthetical references
% ...as shown in recent studies (Simpson et al., 2019)
% \citeA for in-text citations
% ...Simpson et al. (2019) have shown...
%
%
%...as shown by \citeA{jskilby}.
%...as shown by \citeA{lewin76}, \citeA{carson86}, \citeA{bartoldy02}, and \citeA{rinaldi03}.
%...has been shown \cite{jskilbye}.
%...has been shown \cite{lewin76,carson86,bartoldy02,rinaldi03}.
%... \cite <i.e.>[]{lewin76,carson86,bartoldy02,rinaldi03}.
%...has been shown by \cite <e.g.,>[and others]{lewin76}.
%
% apacite uses < > for prenotes and [ ] for postnotes
% DO NOT use other cite commands (e.g., \citet, \citep, \citeyear, \nocite, \citealp, etc.).
%

\end{document}